\pdfoutput=1

\documentclass[11pt,twoside,a4paper,cmspaper,final,collab]{cms-tdr}
\def\svnVersion{257106}\def\cmsCernNo{2013-037}\def\cmsCernDate{\today}\def\cmsMessage{Published in the European Physical Journal C as \href{http://dx.doi.org/10.1140/epjc/s10052-014-3014-0}{\doi{10.1140/epjc/s10052-014-3014-0.}}}

\begin{document}\cmsNoteHeader{TOP-12-018}

\hyphenation{had-ron-i-za-tion}
\hyphenation{cal-or-i-me-ter}
\hyphenation{de-vices}
\RCS$Revision: 253047 $
\RCS$HeadURL: svn+ssh://svn.cern.ch/reps/tdr2/papers/TOP-12-018/trunk/TOP-12-018.tex $
\RCS$Id: TOP-12-018.tex 253047 2014-07-25 21:56:16Z alverson $
\newlength\cmsFigWidth
\ifthenelse{\boolean{cms@external}}{\setlength\cmsFigWidth{0.85\textwidth}}{\setlength\cmsFigWidth{0.6\textwidth}}
\ifthenelse{\boolean{cms@external}}{\providecommand{\cmsLeft}{top}}{\providecommand{\cmsLeft}{left}}
\ifthenelse{\boolean{cms@external}}{\providecommand{\cmsRight}{bottom}}{\providecommand{\cmsRight}{right}}
\ifthenelse{\boolean{cms@external}}{\providecommand{\breakhere}{\linebreak[4]}}{\providecommand{\breakhere}}{\relax}
\cmsNoteHeader{TOP-12-018}
\title{Measurement of jet multiplicity distributions in \texorpdfstring{\ttbar}{t t-bar} production in pp collisions at \texorpdfstring{$\sqrt{s} = 7\TeV$}{sqrt(s) = 7 TeV}}

\date{\today}

\newcommand{\ttjets}{\text{\ttbar+jets}\xspace}
\newcommand{\MADGPYT} {\textsc{MadGraph+pythia}\xspace}
\newcommand{\POWPYT} {{\textsc{powheg+pythia}}\xspace}
\newcommand{\POWHER} {{\textsc{powheg+herwig}}\xspace}
\newcommand{\MCNLOHER} {\textsc{mc@nlo+herwig}\xspace}
\newcommand{\Irel}{\ensuremath{I_\text{rel}}\xspace}
\newcommand{\diel}{\ensuremath{\Pep\Pem}\xspace}

\abstract{
The normalised differential top quark-antiquark production cross section is measured as a function of the jet multiplicity in proton-proton collisions at a centre-of-mass energy of 7\TeV at the LHC with the CMS detector. The measurement is performed in both the dilepton and lepton+jets decay channels using data corresponding to an integrated luminosity of 5.0\fbinv.
Using a procedure to associate jets to decay products of the top quarks, the differential cross section of the \ttbar production is determined as a function of the additional jet multiplicity in the lepton+jets channel. Furthermore, the fraction of events with no additional jets is measured in the dilepton channel, as a function of the threshold on the jet transverse momentum.
The measurements are compared with predictions from perturbative quantum chromodynamics and no significant deviations are observed.}

\hypersetup{%
pdfauthor={CMS Collaboration},%
pdftitle={Measurement of jet multiplicity distributions in t t-bar production in pp collisions at sqrt(s) = 7 TeV},%
pdfsubject={CMS},%
pdfkeywords={CMS, physics, top quark}}

\maketitle

\section{Introduction}
\label{sec:introduction}

Precise measurements of the top quark-antiquark (\ttbar) production cross section and top-quark properties performed at the CERN Large Hadron Collider (LHC)
provide crucial information for testing the predictions of perturbative quantum chromodynamics (QCD) at large energy scales and in processes with multiparticle final states.

About half of the \ttbar events are expected to be accompanied by additional hard jets that do not originate from the decay of the \ttbar pair (\ttjets). In this paper, these jets will be referred to as \textit{additional jets}. These processes typically arise from either initial- or final-state QCD radiation, providing an essential handle to test the validity and completeness of higher-order QCD calculations of processes leading to multijet events. Calculations at next-to-leading order (NLO) are available for \ttbar production in association with one~\cite{Dittmaier:2007wz} or two~\cite{Bevilacqua:2011aa} additional jets. The correct description of \ttjets production is important to the overall LHC physics program since it constitutes an important background to processes with multijet final states,
such as associated Higgs-boson production with a \ttbar pair, with the Higgs boson decaying into a $\bbbar$ pair, or final states predicted in supersymmetric theories. Anomalous production of additional jets accompanying a \ttbar pair could be a sign of new physics beyond the standard model~\cite{bib:topres}.

This paper presents studies of the \ttbar production with additional jets in the final state using data collected in proton-proton (pp) collisions with centre-of-mass energy $\sqrt{s} = 7\TeV$ with the Compact Muon Solenoid (CMS) detector~\cite{bib:CMS}. The analysis uses data recorded in 2011, corresponding to a total integrated luminosity of $5.0 \pm 0.1$\fbinv.
For the first time, the \ttbar cross section is measured differentially as a function of jet multiplicity and characterised both in terms of the total number of jets in the event, as well as the number of additional jets with respect to the leading-order hard-interaction final state.
Kinematic properties of the additional jets are also investigated. The results are corrected for detector effects and compared at particle level with theoretical predictions obtained using different Monte Carlo (MC) event generators.

The differential cross sections as a function of jet multiplicity are measured in both the dilepton ($\Pe\Pe$, $\Pgm\Pgm$, and $\Pe\Pgm$) and $\ell$+jets ($\ell$ = $\Pe$ or $\Pgm$) channels.
For the dilepton channel, data containing two oppositely-charged leptons and at least two jets in the final state are used, while for the $\ell$+jets channel, data containing a single isolated lepton and at least three jets are used.
Following the analysis strategy applied to the measurement of other \ttbar differential cross sections~\cite{bib:TOP-11-013_paper}, the results are normalised to the inclusive cross section measured \textit{in situ}, eliminating systematic uncertainties related to the normalisation. Lastly, the fraction of events that do not contain additional jets (\textit{gap fraction}), first measured by ATLAS~\cite{bib:atlas2}, is determined in the dilepton channel as a function of the threshold on the transverse momentum (\pt) of the leading additional jet and of the scalar sum of the \pt of all additional jets.

The measurements are performed in the visible phase space, defined as the kinematic region in which all selected final-state objects are produced within the detector acceptance. This avoids additional model uncertainties due to the extrapolation of the measurements into experimentally inaccessible regions of phase space.

The paper is structured as follows. A brief description of the CMS detector is provided in Sect.~\ref{sec:detector}. Section~\ref{sec:theory} gives a description of the event simulation, followed by details of the object reconstruction and event selection in Sect.~\ref{sec:selection}. A discussion of the sources of systematic uncertainties is given in Sect.~\ref{sec:syst}. The measurement of the differential cross section is presented as a function of the jet multiplicity in Sect.~\ref{sec:results} and as a function of the additional jet multiplicity in Sect.~\ref{sec:addJets}. The study of the additional jet gap fraction is described in Sect.~\ref{sec:gap}. Finally, a summary is given in Sect.~\ref{sec:summary}.

\section{The CMS detector}\label{sec:detector}
The central feature of the CMS apparatus is a superconducting solenoid, 13\unit{m} in length and 6\unit{m} in diameter, which provides an axial magnetic field of 3.8\unit{T}. The bore of the solenoid is outfitted with various particle detection systems. Charged-particle trajectories are measured with silicon pixel and strip trackers, covering $0 \leq \phi < 2\pi$ in azimuth and $\abs{\eta}<2.5$ in pseudorapidity, where $\eta$ is defined as $ \eta =- \ln [\tan (\theta/2)]$ , with $\theta$
being the polar angle of the trajectory of the particle with respect to the anticlockwise-beam direction. A lead tungstate
crystal electromagnetic calorimeter (ECAL) and a brass/scintillator hadron calorimeter (HCAL) surround the tracking volume. The calorimetry provides excellent resolution in energy for electrons and hadrons within $\abs{\eta}<3.0$. Muons are measured up to $\abs{\eta}<2.4$ using gas-ionisation detectors embedded in the steel flux return yoke outside the solenoid. The detector is nearly hermetic, providing accurate measurements of any imbalance in momentum in
the plane transverse to the beam direction. The two-level trigger system selects most interesting final states for further analysis.
A detailed description of the CMS detector can be found in Ref.~\cite{bib:CMS}.

\section{Event simulation}
\label{sec:theory}

The reference simulated \ttbar sample used in the analysis is generated with the \MADGRAPH (v.~5.1.1.0) matrix element generator~\cite{Alwall:2011uj}, with up to three additional partons. The generated events are subsequently processed using \PYTHIA (v.~6.424)~\cite{Sjostrand:2006za} to add parton showering using the MLM prescription~\cite{Mangano:2006rw} for removing the overlap in phase space between the matrix element and the parton shower approaches. The \PYTHIA Z2 tune is used to describe the underlying event~\cite{UEPAS}.
The top-quark mass is assumed to be $m_{\cPqt}$=172.5\GeV. The proton structure is described by the CTEQ6L1~\cite{bib:cteq} parton distribution functions (PDFs).

The \MADGRAPH generator is used to simulate W+jets and $\cPZ/\gamma^*$+jets production. Single-top-quark events ($s$-, $t$-, and tW-channels) are simulated using \POWHEG (r1380)~\cite{Nason:2004rx, Frixione:2007vw,Alioli:2009je,Alioli:2009je,Re:2010bp}. Diboson (WW, WZ, and ZZ) and QCD multijet events are simulated using \PYTHIA.

Additional \ttbar and W+jets \MADGRAPH samples are generated using different choices for the common factorisation and renormalisation scale ($\mu_F^2=\mu_R^2=Q^2$) and for the jet-parton matching threshold. These are used to determine the systematic uncertainties due to model uncertainties and for comparisons with the measured distributions.
The nominal $Q^2$ scale is defined as $m_{\cPqt}^2 +\sum{\pt^2(\text{jet})}$. This is varied between $4Q^2$ and $Q^2/4$. For the reference \MADGRAPH sample, a jet-parton matching threshold of 20\GeV is chosen, while for the up and down variations, thresholds of 40 and 10\GeV are used, respectively.

In addition to \MADGRAPH, samples of \ttbar events are generated with \POWHEG and \MCATNLO (v.~3.41)~\cite{bib:mcatnlo}.
The CTEQ6M~\cite{bib:cteq} PDF set is used in both cases. Both \POWHEG and \MCATNLO match calculations to full NLO accuracy with parton shower MC generators. For \POWHEG, \PYTHIA is chosen for hadronisation and parton shower simulation, with the same Z2 tune utilised for other samples. For \MCATNLO, \HERWIG (v.~6.520)~\cite{bib:herwig} with the default tune is used.

{\tolerance=800
For comparison with the measured distributions, the event yields in the simulated samples are normalised to an integrated luminosity of 5.0\fbinv according to their theoretical cross sections. These are taken from next-to-next-to-leading-order (NNLO) (W+jets and $\cPZ/\gamma^*$+jets), NLO plus next-to-next-to-leading-log (NNLL) (single-top-quark $s$-~\cite{bib:schan}, $t$-~\cite{bib:tchan} and tW-channels~\cite{bib:twchan}), NLO (diboson~\cite{bib:mcfm:diboson}), and leading-order (LO) (QCD multijet~\cite{Sjostrand:2006za}) calculations.
For the simulated \ttbar sample, the full NNLO+NNLL calculation, performed with the \textsc{Top++} 2.0 program~\cite{Czakon:2011xx}, is used. The PDF and $\alpha_S$ uncertainties are estimated using the PDF4LHC prescription~\cite{Alekhin:2011sk, Botje:2011sn} with the \breakhere MSTW2008nnlo68cl~\cite{Martin:2009iq}, CT10 NNLO~\cite{Lai:2010vv, Gao:2013xoa}, and \breakhere NNPDF2.3 5f FFN~\cite{Ball:2012cx} PDF sets, and added in quadrature to the scale uncertainty to obtain a \ttbar production cross section of $177.3^{+10.1}_{-10.8}$\unit{pb} (for a top-quark mass value of 172.5\GeV).
\par}

All generated samples are passed through a full detector simulation using \GEANTfour~\cite{Agostinelli:2002hh}, and the number of additional pp collisions (pileup) is matched to the real distribution as inferred from data.

\section{Event reconstruction and selection}
\label{sec:selection}

The event selection is based on the reconstruction of the \ttbar decay products. The top quark decays almost exclusively into a W boson and a b quark. Only the subsequent decays of one or both W bosons to a charged lepton and a neutrino are considered here.
Candidate events are required to contain the corresponding reconstructed objects:
isolated leptons and jets. The requirement of the presence of jets associated with b quarks or antiquarks (b jets) is used to increase the purity of the selected sample. The selection has been optimised independently in each channel to maximise the signal content and background rejection.

\subsection{Lepton, jet, and missing transverse energy reconstruction}
\label{subsec:reconstruction}

Events are reconstructed using a particle-flow (PF) technique~\cite{CMS-PAS-PFT-09-001, CMS-PAS-PFT-10-002}, in which signals from all CMS sub-detectors are combined to identify and reconstruct the individual particle candidates produced in the pp collision. The reconstructed particles include muons, electrons, photons, charged hadrons, and neutral hadrons. Charged particles are required to originate from the primary collision vertex, defined as the vertex with the highest sum of transverse momenta of all reconstructed tracks associated to it. Therefore, charged hadron candidates from pileup events, \ie originating from a vertex other than the one of the hard interaction, are removed before jet clustering on an event-by-event basis. Subsequently, the remaining neutral-hadron pileup component is subtracted at the level of jet energy correction~\cite{bib:neutrals}.

Electron candidates are reconstructed from a combination of their track and energy deposition in the ECAL~\cite{CMS-PAS-EGM-10-004}.
In the dilepton channel, they are required to have a transverse momentum $\pt>20\GeV$, while in the $\ell$+jets channel they are required to have $\pt>30\GeV$. In both cases they are required to be reconstructed within $\abs{\eta}<2.4$, and electrons from identified photon conversions are rejected. As an additional quality criterion, a relative isolation variable \Irel is computed. This is defined as the sum of the \pt of all neutral and charged reconstructed PF candidates inside a cone around the lepton (excluding the lepton itself) in the $\eta$-$\phi$ plane with radius $\Delta R\equiv\sqrt{\smash[b]{(\Delta\eta)^2+(\Delta\phi)^2}}<0.3$, divided by the \pt of the lepton. In the dilepton (e+jets) channel, electrons are selected as isolated if $\Irel<0.12~(0.10)$.

Muon candidates are reconstructed from tracks that can be matched between the silicon tracker and the muon system~\cite{muon-performance}. They are required to have a transverse momentum $\pt>20\GeV$ within the pseudorapidity interval $\abs{\eta}<2.4$ in the dilepton channel, and to have $\pt>30 \GeV$ and $\abs{\eta}<2.1$ in the $\ell$+jets channel. Isolated muon candidates are selected by demanding a relative isolation of $\Irel<0.20\,(0.125)$ in the dilepton ($\mu$+jets) channel.

Jets are reconstructed by clustering the particle-flow candidates~\cite{bib:JME-10-011:JES} using the anti-\kt algorithm with a distance parameter of $0.5$~\cite{Cacciari:2005hq,Cacciari:2008gp}. An offset correction is applied to take into account the extra energy clustered in jets due to pileup, using the FastJet algorithm~\cite{Cacciari:2011ma} based on average pileup energy density in the event.
The raw jet energies are corrected to establish a relative uniform response of the calorimeter in $\eta$ and a calibrated absolute response in \pt. Jet energy corrections are derived from the simulation, and are confirmed with \textit{in situ} measurements with the energy balance of dijet and photon+jet events~\cite{bib:JME-10-011:JES}.
Jets are selected within $\abs{\eta}<2.4$ and with $\pt >30\,(35)\GeV$ in the dilepton ($\ell$+jets) channel.

Jets originating from b quarks or antiquarks are identified with the Combined Secondary Vertex algorithm~\cite{btv}, which provides a b-tagging discriminant by combining secondary vertices and track-based lifetime information.
The chosen working point used in the dilepton channel corresponds to an efficiency for tagging a b jet of about 80--85\%, while the probability to misidentify light-flavour or gluon jets as b jets (mistag rate) is around 10\%. In the $\ell$+jets channel, a tighter requirement is applied, corresponding to a b-tagging efficiency of about 65--70\% with a mistag rate of 1\%. The probability to misidentify a c jet as b jet is about 40\% and 20\% for the working points used in the dilepton and $\ell$+jets channels respectively~\cite{btv}.

The missing transverse energy (\MET) is defined as the magnitude of the sum of the momenta of all reconstructed PF candidates in the plane transverse to the beams.

\subsection{Event selection}
\label{subsec:selection}

Dilepton events are collected using combinations of triggers which require two leptons fulfilling \pt and isolation criteria.
During reconstruction, events are selected if they contain at least two isolated leptons (electrons or muons) of opposite charge and at least two jets, of which at least one is identified as a b jet. Events with a lepton pair invariant mass smaller than 12\GeV are removed in order to suppress events from heavy-flavour resonance decays. In the $\Pe\Pe$ and $\Pgm\Pgm$ channels, the dilepton invariant mass is required to be outside a Z-boson mass window of $91\pm15\GeV$ (\textit{Z-boson veto}), and \MET is required to be larger than 30\GeV.

A kinematic reconstruction method~\cite{bib:TOP-11-013_paper} is used to determine the kinematic properties of the \ttbar pair and to identify the two b jets originating from the decay of the top quark and antiquark. In the kinematic reconstruction the following constraints are imposed: the \MET originated entirely from the two neutrinos; the reconstructed W-boson invariant mass of 80.4\GeV~\cite{pdg} and the equality of the reconstructed top quark and antiquark masses. The remaining ambiguities are resolved by prioritising those event solutions with two or one b-tagged jets over solutions using untagged jets.
Finally, among the physical solutions, the solutions are ranked according to how the neutrino energies match with a simulated neutrino energy spectrum and the highest ranked one is chosen. The kinematic reconstruction yields no valid solution for about 11\% of the events. These are excluded from further analysis. A possible bias due to rejected solutions has been studied and found to be negligible.

In the e+jets channel, events are triggered by an isolated electron with $\pt>25$\GeV and at least three jets with $\pt> 30\GeV$. Events in the $\mu+$jets channel are triggered by the presence of an isolated muon with $\pt>24\GeV$ fulfilling $\eta$ requirements. Only triggered events that have exactly one high-\pt isolated lepton are retained in the analysis. In the e+jets channel, events are rejected if any additional electron is found with $\pt > 20\GeV$, $\abs{\eta}<2.5$, and relative isolation $\Irel<0.20$. In the $\mu$+jets channel, events are rejected if any electron candidate with $\pt >15\GeV$, $\abs{\eta}<2.5$ and $\Irel<0.20$ is reconstructed. In both $\ell$+jets channels events with additional muons with $\pt > 10\GeV$, $\abs{\eta}<2.5$, and relative isolation $\Irel<0.20$ are rejected.
The presence of at least three reconstructed jets is required. At least two of them are required to be b-tagged.

Only \ttbar events from the decay channel under study are considered as signal. All other \ttbar events are considered as background, including those containing leptons from $\tau$ decays, which are the dominant contribution to this background.

\subsection{Background estimation}
\label{subsec:bg}

After the full event selection is applied, the dominant background in the e$\mu$ channel comes from other \ttbar decay modes, estimated using simulation. In the $\Pe\Pe$ and $\Pgm\Pgm$ channels, it arises from $\cPZ/\gamma^*$+jets production. The normalisation of this backround contribution is derived from data using the events rejected by the Z-boson veto,
scaled by the ratio of events failing and passing this selection estimated in simulation ($R_{\text{out/in}}$)~\cite{Chatrchyan:2011nb}.
The number of $\cPZ/\gamma^*\text{+jets}\to\Pe\Pe /\Pgm\Pgm$ events near the Z-boson peak, $N_{\cPZ/\gamma^*}^\text{in}$, is given by the number of all
events failing the Z-boson veto, $N^\text{in}$, after subtracting the contamination from non-$\cPZ/\gamma^*$+jets processes. This contribution is extracted from e$\mu$ events passing the same selection, $N_{\Pe\Pgm}^\text{in}$,  and corrected for the differences between the electron and muon identification efficiencies using a correction factor $k$.
The $\cPZ/\gamma^*$+jets contribution is thus given by
\begin{equation}
N^{\text{out}} = R_{\text{out/in}} N_{\cPZ/\gamma^*}^\text{in}  =  R_{\text{out/in}} (N^\text{in} - 0.5 k N_{\Pe\Pgm}^\text{in})
\end{equation}
The factor $k$ is estimated from $k^2=N_{\Pe\Pgm}/N_{\Pe\Pe}$ $(N_{\Pe\Pgm}/N_{\Pgm\Pgm}$) for the $\cPZ/\gamma^*\to\diel\,(\Pgmp\Pgmm)\text{+jets}$ contribution, respectively. Here $N_{\Pe\Pe}$ ($N_{\Pgm\Pgm}$) is the number of $\Pe\Pe$ ($\Pgm\Pgm$) events in the Z-boson region, without the requirement on \MET. The remaining backgrounds, including single-top-quark, W+jets, diboson, and QCD multijet events are estimated from simulation.

In the $\ell$+jets channel, the main background contributions arise from W+jets and QCD multijet events, which are greatly suppressed by the b-tagging requirement. A procedure based on control samples in data is used to extract the QCD multijet background. The leptons in QCD multijet events are expected to be less isolated than leptons from other processes. Thus, inverting the selection on the lepton relative isolation provides a relatively pure sample of QCD multijet events in data. Events passing the standard event selection but with an $\Irel$ between 0.3 and 1.0, and with at least one b-tagged jet are selected. The sample is divided in two: the sideband region (one b jet) and the signal region ($\geq$2 b jets). The shape of the QCD multijet background is taken from the signal region, and the normalisation is determined from the sideband region. In the sideband region, the \MET distribution of  the QCD multijet model,  other sources of background (determined from simulation), and the \ttbar signal are fitted to data. The resulting scaling of QCD multijet background is applied to the QCD multijet shape from the signal region.

Since the initial state of LHC collision is enriched in up quarks with respect to down quarks, more W bosons are produced
with positive charge than negative charge. In leptonic W-boson decays, this translates into a lepton charge asymmetry $\mathcal{A}$. Therefore, a difference between the number of events with a positively charged lepton and those with a negatively charged lepton ($\Delta\pm$) is observed.
In data, this quantity ($\Delta\pm^\text{data}$) is proportional to the number of W+jets events when assuming that only the charge asymmetry from W-boson production is significant.
The charge asymmetry has been measured by CMS~\cite{bib:Wasym} and found to be well described by the simulation, thus the simulated value
can be used to extract the number of W+jets events from data: $N_{\PW\text{+jets}}^{\text{data}}=\Delta\pm^\text{data}/\mathcal{A}$.
The correction factor on the W+jets normalisation, calculated before any b-tagging requirement, is between 0.81 and 0.92 depending on the W decay channel and the jet selection.
Subsequently, b-tagging is applied to obtain the number of W+jets events in the signal region.

In addition, a heavy-flavour correction must be applied on the W+jets sample to account for the differences observed between data and simulation~\cite{CMS-PAS-TOP-11-003}. Using the matching between selected jets and generated partons, simulated events are classified as containing at least one b jet (W+bX), at least one c jet and no b jets (W+cX), or containing neither b jets nor c jets (W+light quarks). The rate of W+bX events is multiplied by $2\pm1$ and the rate of W+cX events is multiplied by $1^{+1.0}_{-0.5}$. No correction is applied to W+light-jets events. These correction factors are calculated in~\cite{CMS-PAS-TOP-11-003} in a phase space which is close to the one used in the analysis. The uncertainties in the correction factors are taken into account as systematic uncertainties.
The total number of W+jets events is modified to conserve this number when applying the heavy-flavour corrections. The remaining backgrounds, originating from single-top-quark, diboson, and $\cPZ/\gamma^*$+jets processes, are small and their contributions are estimated using simulation.

\begin{figure*}[htb!]
  \centering
      \includegraphics[width=0.49 \textwidth]{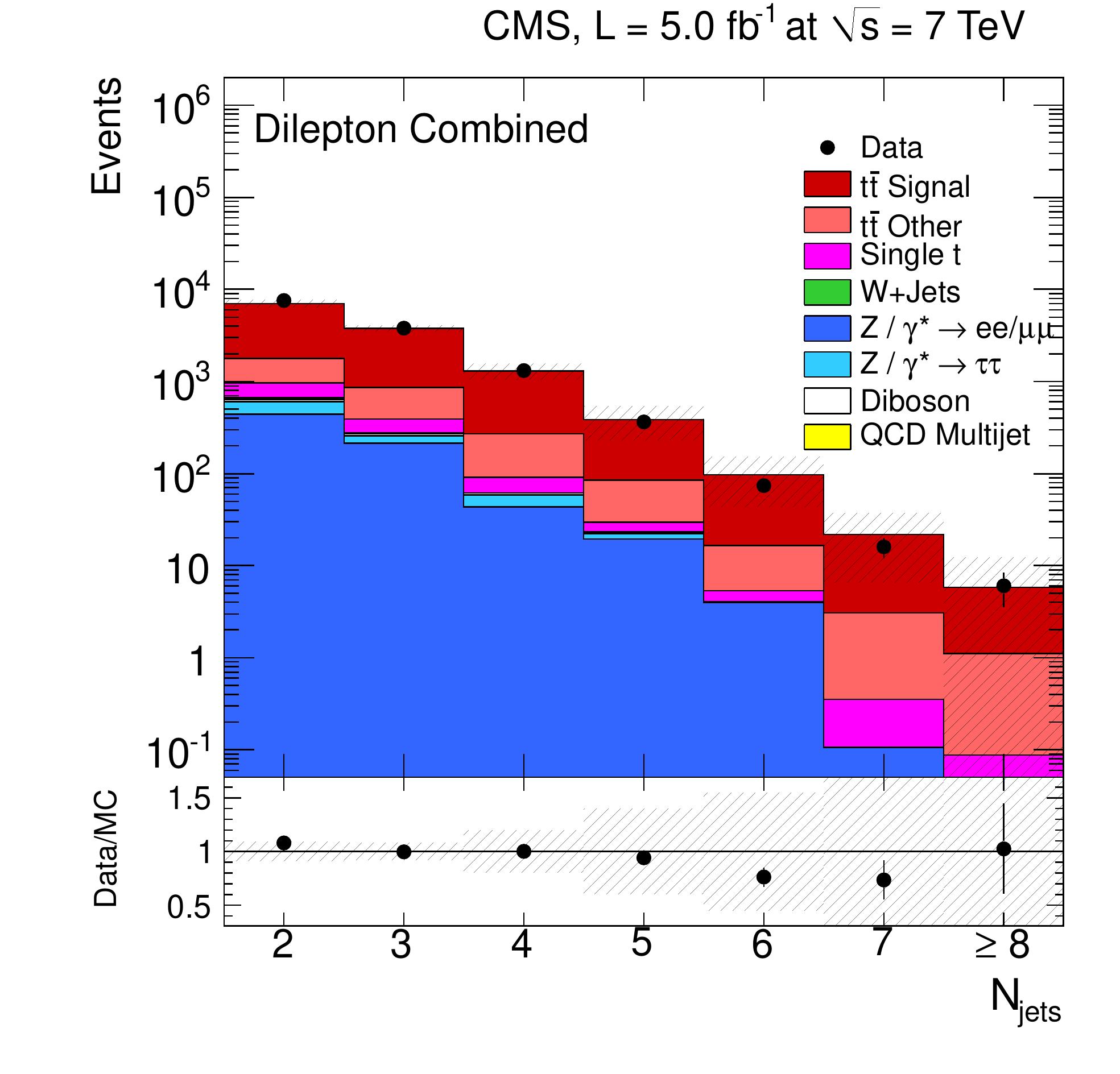}
      \includegraphics[width=0.49 \textwidth]{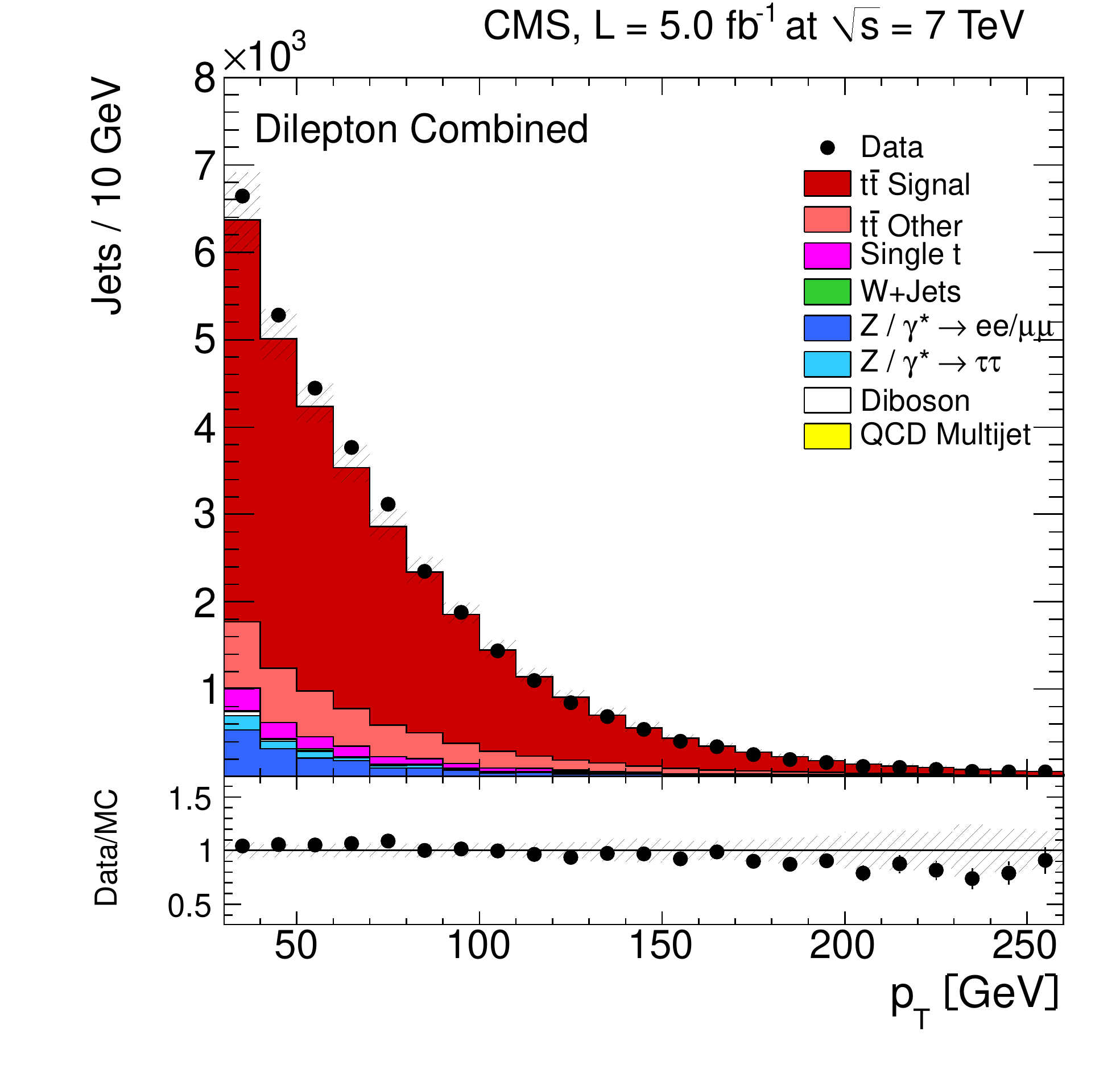}
      \includegraphics[width=0.49 \textwidth]{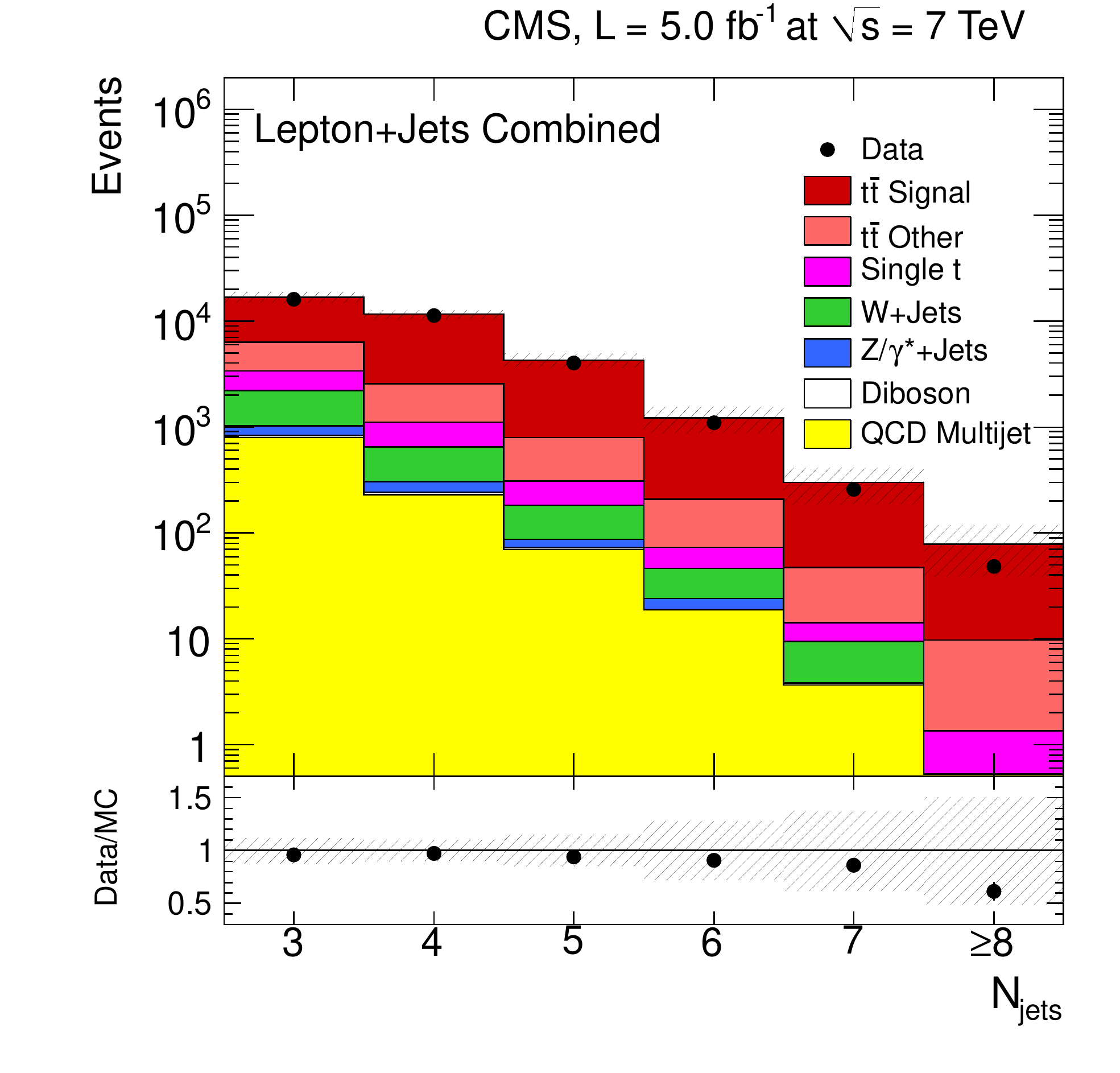}
      \includegraphics[width=0.49 \textwidth]{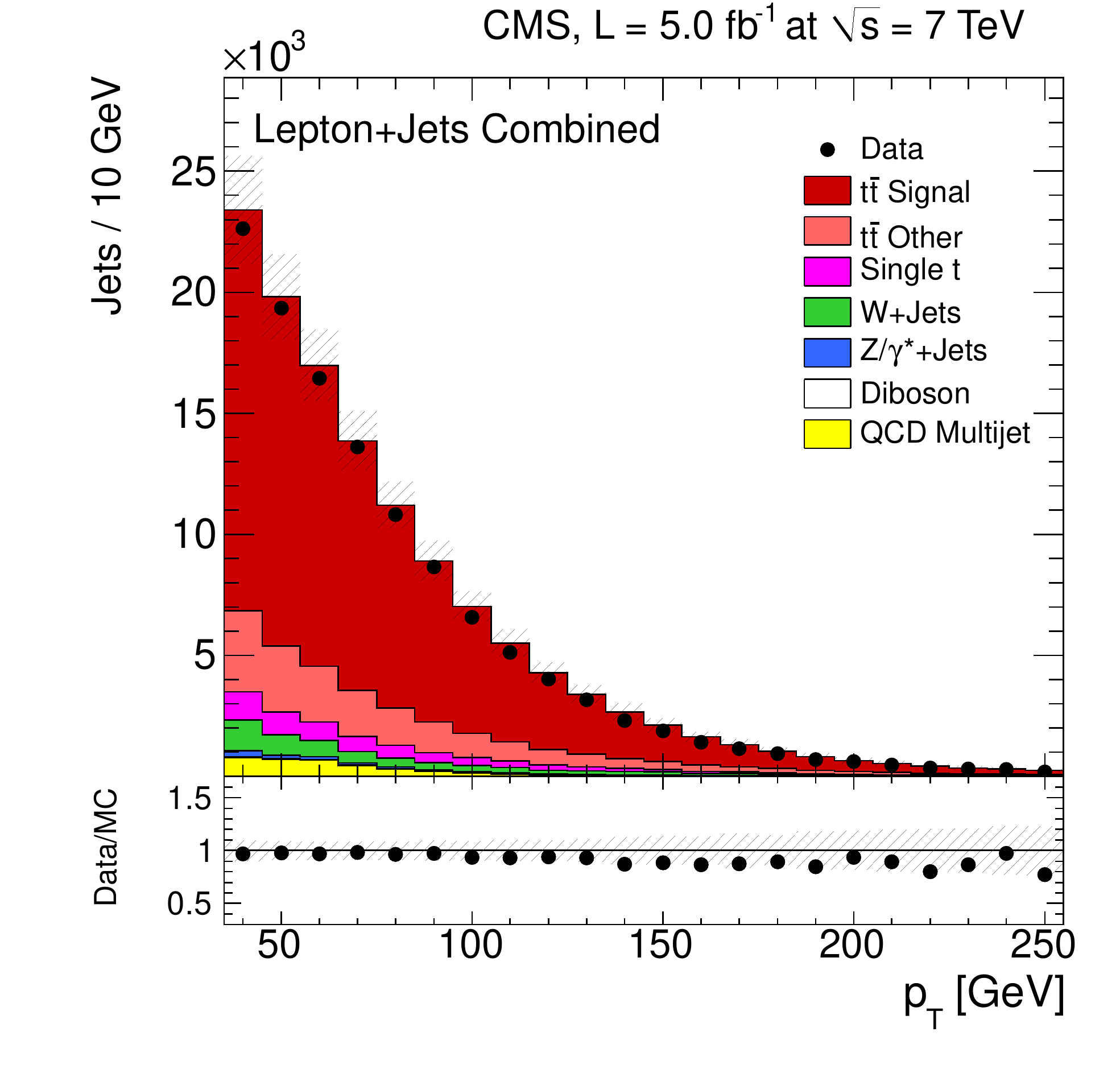}
\caption{Number of reconstructed jets (left) and jet \pt spectrum (right) after event selection in the dilepton channel for jets with $\pt>30$\GeV (top), and in the $\ell$+jets channel for jets with $\pt>35$\GeV (bottom). The hatched band represents the combined effect of all sources of systematic uncertainty.}
\label{fig:NJets}
\end{figure*}

The multiplicity and the \pt distributions of the selected reconstructed jets are shown for the dilepton and $\ell$+jets channels in Fig.~\ref{fig:NJets}.
Good agreement for the jet multiplicity is observed between data and simulation for up to 5 (6) jets in the dilepton ($\ell$+jets) channels. For higher jet multiplicities, the simulation predicts slightly more events than observed in data. The modelling of the jet \pt spectrum in data is shifted towards smaller values, covered by the systematic uncertainties. The uncertainty from all systematic sources, which are described in Sect.~\ref{sec:syst}, is determined by estimating their effect on both the normalisation and the shape. The size of these global uncertainties does not reflect those in the final measurements, since they are normalised and, therefore, only affected by shape uncertainties.

\section{Systematic uncertainties}
\label{sec:syst}
Systematic uncertainties in the measurement arise from detector effects, background modelling, and theoretical assumptions. Each systematic uncertainty is investigated separately and estimated for each bin of the measurement by varying the corresponding efficiency, resolution, or scale within its uncertainty. For each variation, the measured normalised differential cross section is recalculated, and the difference between the varied result and the nominal result in each bin is taken as systematic uncertainty. The overall uncertainty in the measurement is obtained by adding all contributions in quadrature. The sources of systematic uncertainty, described below, are assumed to be uncorrelated.

\begin{itemize}

\item {\textbf{Jet energy}}
The impact of the jet energy scale (JES)~\cite{bib:JME-10-011:JES} is determined by varying the \pt of all jets by the JES uncertainty, which is typically below 3\%. The uncertainty due to the jet energy resolution (JER)~\cite{CMS-PAS-JME-10-014} is estimated by varying the nominal value by $\pm 1\sigma$. 

\item {\textbf{\ttbar model uncertainties}}
Uncertainties originating from theoretical assumptions on the renormalisation and factorisation scales, the jet-parton matching threshold, the hadronisation model, and the colour reconnection modelling~\cite{colorrec}, are determined by repeating the analysis, replacing the reference \MADGRAPH signal simulation by other simulation samples. In particular, the impact of the former sources is assessed with \MADGRAPH samples with the renormalisation and factorisation scales simultaneously varied from the nominal $Q^2$ values to $4 Q^2$ and $Q^2/4$ and with jet-parton matching threshold varied to 40 and 10\GeV. The uncertainties from ambiguities in modeling colour reconnection effects are estimated by comparing simulations of an underlying event tune including colour reconnection to a tune without it (the Perugia 2011 and Perugia 2011 noCR tunes described in~\cite{Skands:2010ak}). The hadronisation model uncertainty is estimated by comparing samples simulated with \POWHEG and \MCATNLO, using \PYTHIA and \HERWIG, respectively, for hadronisation.
The uncertainty arising from the PDFs is assessed by reweighting the \ttbar signal sample according to the 44 CTEQ66 error PDF sets, at 90\% confidence level. The effects of these variations are added in quadrature.

\item {\textbf{Background}} The uncertainty due to the normalisation of the backgrounds that are taken from simulation is determined by varying the cross section by $\pm$30\%~\cite{bib:ttxsljets,bib:top11005}. This takes into account the uncertainty in the predicted cross section and all other sources of systematic uncertainty.

In the dilepton channels, the contribution from \breakhere$\cPZ/\gamma^*\text{+jets}$ processes as determined from data is varied in normalisation by $\pm$30\%~\cite{Chatrchyan:2011nb}.

In the $\ell$+jets channels, the uncertainty in the W+jets background arises from the contamination of other processes with a lepton charge asymmetry when extracting the rate from data, and from the uncertainty in the heavy-flavour correction factors.
The rate uncertainty is estimated to range from 10\% to 20\%, depending on the channel.
The model uncertainty is estimated using samples with varied renormalisation and factorisation scales and jet-parton matching threshold.

The QCD multijet background modelling uncertainty arises from the choice of the relative isolation requirement on the anti-isolated lepton used for the extraction of the background from data, the influence of the contamination from other processes on the shape, and the extrapolation from the sideband to the signal region. The total uncertainty is about 15\% to more than 100\%, depending on the channel.

\item {\textbf{Other systematic uncertainties}}
The uncertainty associated with the pileup model is determined by varying the minimum bias cross section within its uncertainty of $\pm$8\%. Other uncertainties taken into account originate from lepton trigger, isolation, and identification efficiencies; b-jet tagging efficiency and misidentification probability; integrated luminosity~\cite{CMS-PAS-SMP-12-008}; and the kinematic reconstruction algorithm used in the dilepton channels.

\end{itemize}

In the dilepton channels, the total systematic uncertainty is about 3\% at low jet multiplicities, and increases to about 20\% in the bins with at least five jets. In the $\ell$+jets channels, the total systematic uncertainty is about 6\% at the lowest jet multiplicity, and increases to 34\% for events with at least 8 jets.

The dominant systematic uncertainties for both dilepton and $\ell$+jets channels arise from the JES (with typical values from 2 to 20\%, depending on the jet multiplicity bin and cross section measurement) and the signal model including hadronisation, renormalisation and factorisation scales and jet-parton matching threshold (from 3 to 30\%).
The typical systematic uncertainty due to JER ranges from 0.2 to 3\%,  b-tagging  from 0.3 to 2\%,  pileup from 0.1 to 1.4\%, and background normalisation from 1.6 to 3.8\%. The uncertainty from other sources is below 0.5\%.
The remaining uncertainties on the model arise from PDF and colour reconnection, varying from 0.1 to 1.5\% and from 1 to 5.8\%, respectively.
In all channels, the systematic uncertainty for larger jet multiplicities is dominated by the statistical uncertainty of the simulated samples that are used for the evaluation of modelling uncertainties.

\section{Normalised differential cross section as a function of jet multiplicity}
\label{sec:results}
The differential \ttbar production cross section as a function of the jet multiplicity is measured from the number of signal events after background subtraction and correction for the detector efficiencies and acceptances. The estimated number of background events arising from processes other that \ttbar production ($N_{\text{non \ttbar BG}}$) is directly subtracted from the number of events in data ($N$). The contribution from other \ttbar decay modes is taken into account by correcting $N$--$N_{\text{non \ttbar BG}} $ with the signal fraction, defined as the ratio of the number of selected \ttbar signal events to the total number of selected \ttbar events. This avoids the dependence on the inclusive \ttbar cross section used for normalisation. The normalised differential cross section is derived by scaling to the total integrated luminosity and by dividing the corrected number of events by the cross section measured \textit{in situ} for the same phase space. Because of the normalisation, those systematic uncertainties that are correlated across all bins of the measurement, and therefore only affect the normalisation, cancel out. In order to avoid additional uncertainties due to the extrapolation of the measurement outside of the phase space region probed experimentally, the differential cross section is determined in a visible phase space defined at the particle level by the kinematic and geometrical acceptance of the final-state leptons and jets.

The visible phase space at particle level is defined as follows. The charged leptons from the \ttbar decays are selected with $ \abs{\eta} < 2.4$ in dilepton events and $ \abs{\eta} < 2.5\,(2.1)$ in \Pe+jets ($\mu$+jets) final states, $\pt>20\,(30)\GeV$ in the dilepton ($\ell$+jets) channels. A jet is defined at the particle level in a similar way as described in Sect.~\ref{sec:selection} for the reconstructed jets, by applying the anti-\kt clustering algorithm to all stable particles (including neutrinos not coming from the hard interaction). Particle-level jets are rejected if the selected leptons are within a cone of $\Delta R=0.4$ with respect to the jet, to avoid counting leptons misidentified as jets. A jet is defined as a b jet if it contains the decay products of a b hadron. The two b jets from the \ttbar decay have to fulfill the kinematic requirements $\abs{\eta} < 2.4$ and $\pt>30\,(35)\GeV$ in the dilepton ($\ell$+jets) events. In the $\ell$+jets channels, a third jet with the same properties is also required.

Effects from trigger and detector efficiencies and resolutions, leading to migrations of events across bin boundaries and statistical correlations among neighbouring bins, are corrected by using a regularised unfolding method~\cite{bib:svd, bib:blobel, bib:TOP-11-013_paper}. A response matrix that accounts for migrations and efficiencies is calculated from simulated \ttbar\ events using the reference \MADGRAPH sample. The event migration in each bin is controlled by the purity (number of events reconstructed and generated in one bin divided by the total number of reconstructed events in that bin) and the stability (number of events reconstructed and generated in one bin divided by the total number of generated events in that bin). In these measurements, the purity and stability in the bins is typically 60\% or higher. The generalised inverse of the response matrix is used to obtain the unfolded distribution from the measured distribution by applying a $\chi^2$ technique. To avoid non-physical fluctuations, a smoothing prescription (regularisation) is applied~\cite{bib:james, bib:TOP-11-013_paper}. The unfolded data are subsequently corrected to take into account the acceptance in the particle level phase space.

The measured normalised differential cross sections are consistent among the different dilepton and $\ell$+jets channels. The final results in the dilepton and $\ell$+jets channels are obtained from the weighted average of the individual measurements, using the statistical uncertainty as the weight. 
The result from the combination of e+jets and $\mu$+jets channels is defined for the pseudorapidity range $\abs{\eta} < 2.1$, \ie according to the selection criterion of the $\mu$+jets channel. The difference of this result to that for the pseudorapidity range $\abs{\eta} < 2.5$ has been estimated to be less than 0.4\% in any of the bins of the jet multiplicity distribution. In the combination, the differences in the $\abs{\eta}$-range between $\mu$+jets and e+jets channels are therefore neglected.

\begin{figure*}[htb!]
  \centering
     \includegraphics[width=0.49 \textwidth]{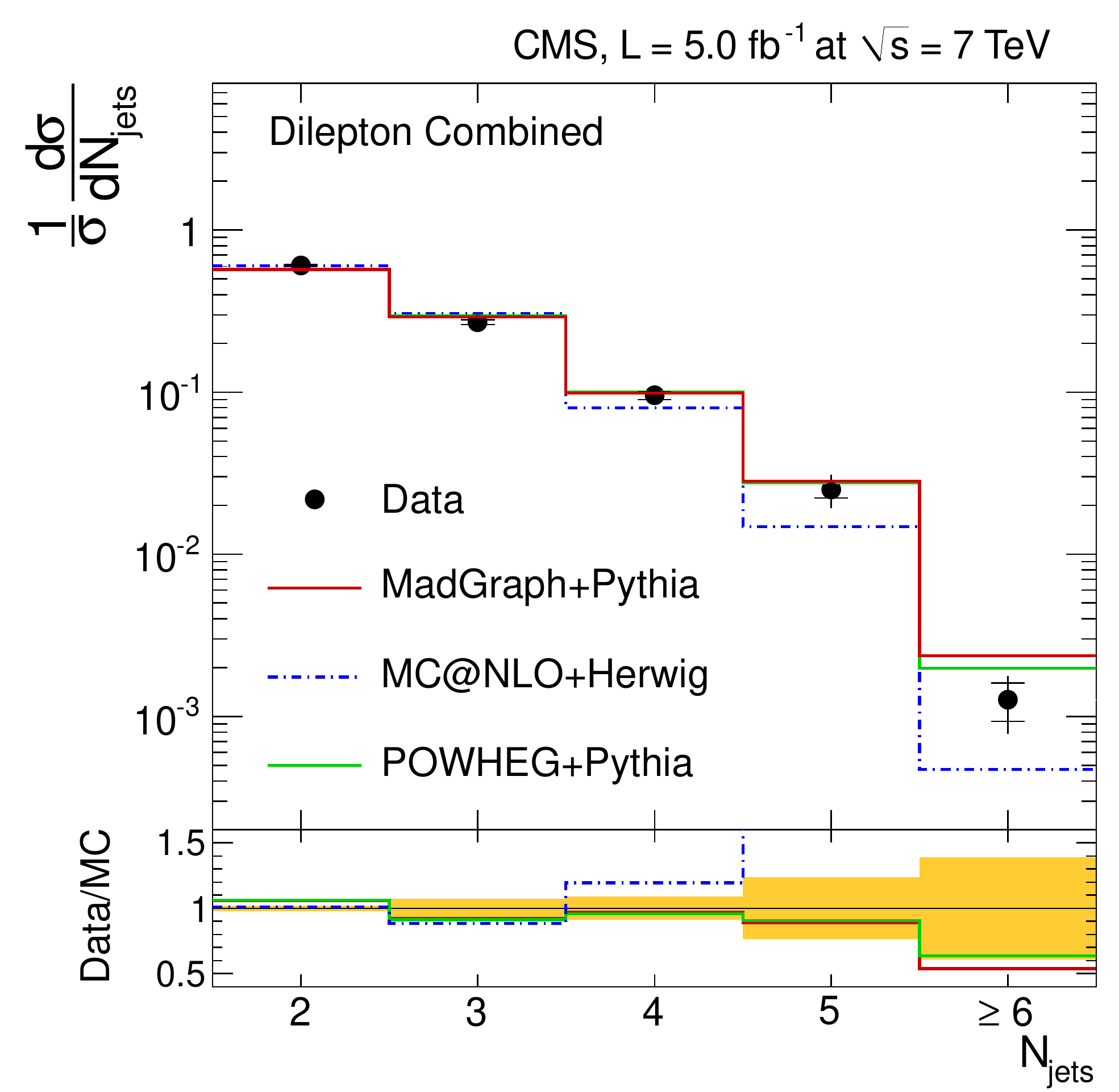}
     \includegraphics[width=0.49 \textwidth]{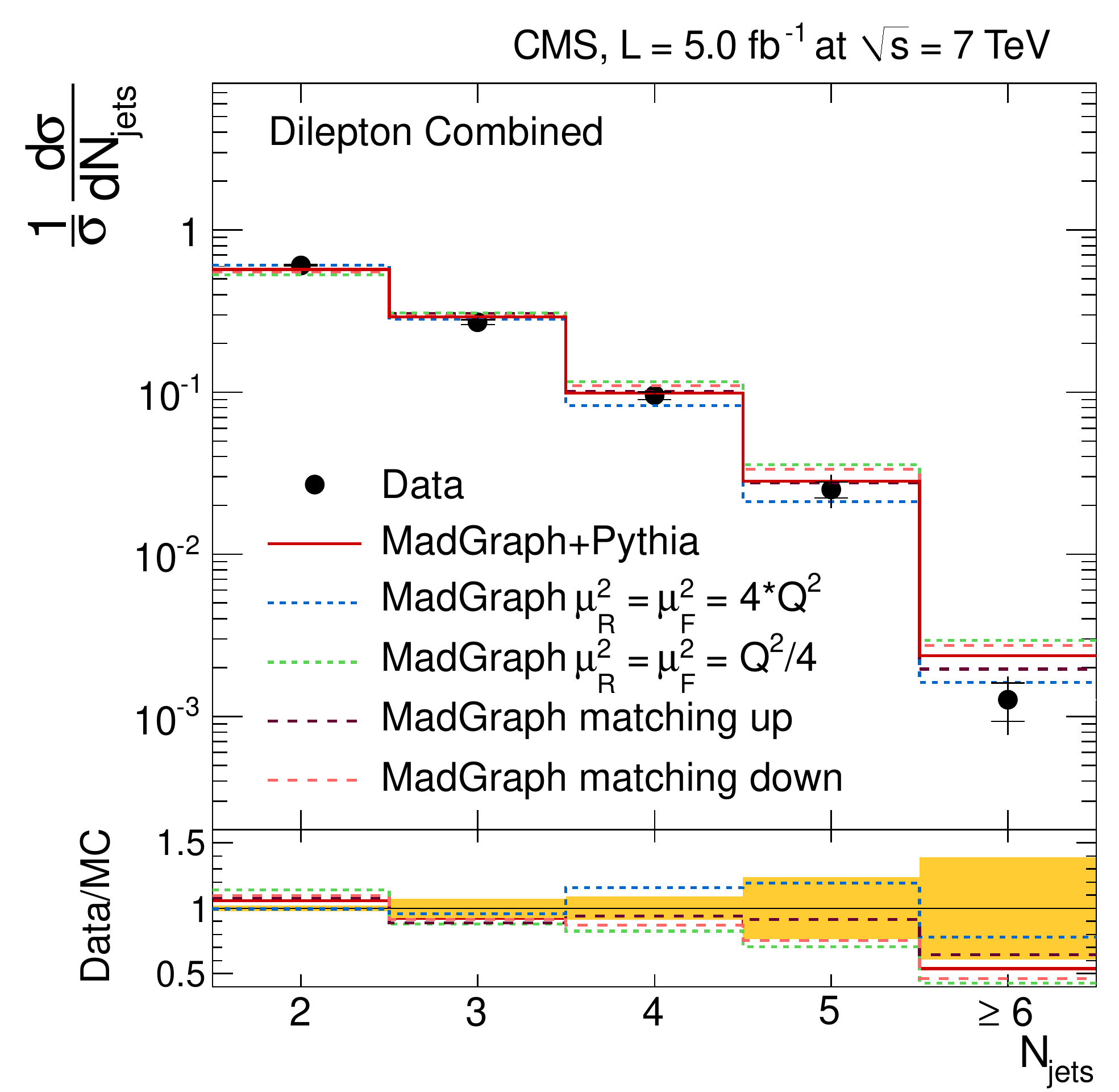}
     \includegraphics[width=0.49 \textwidth]{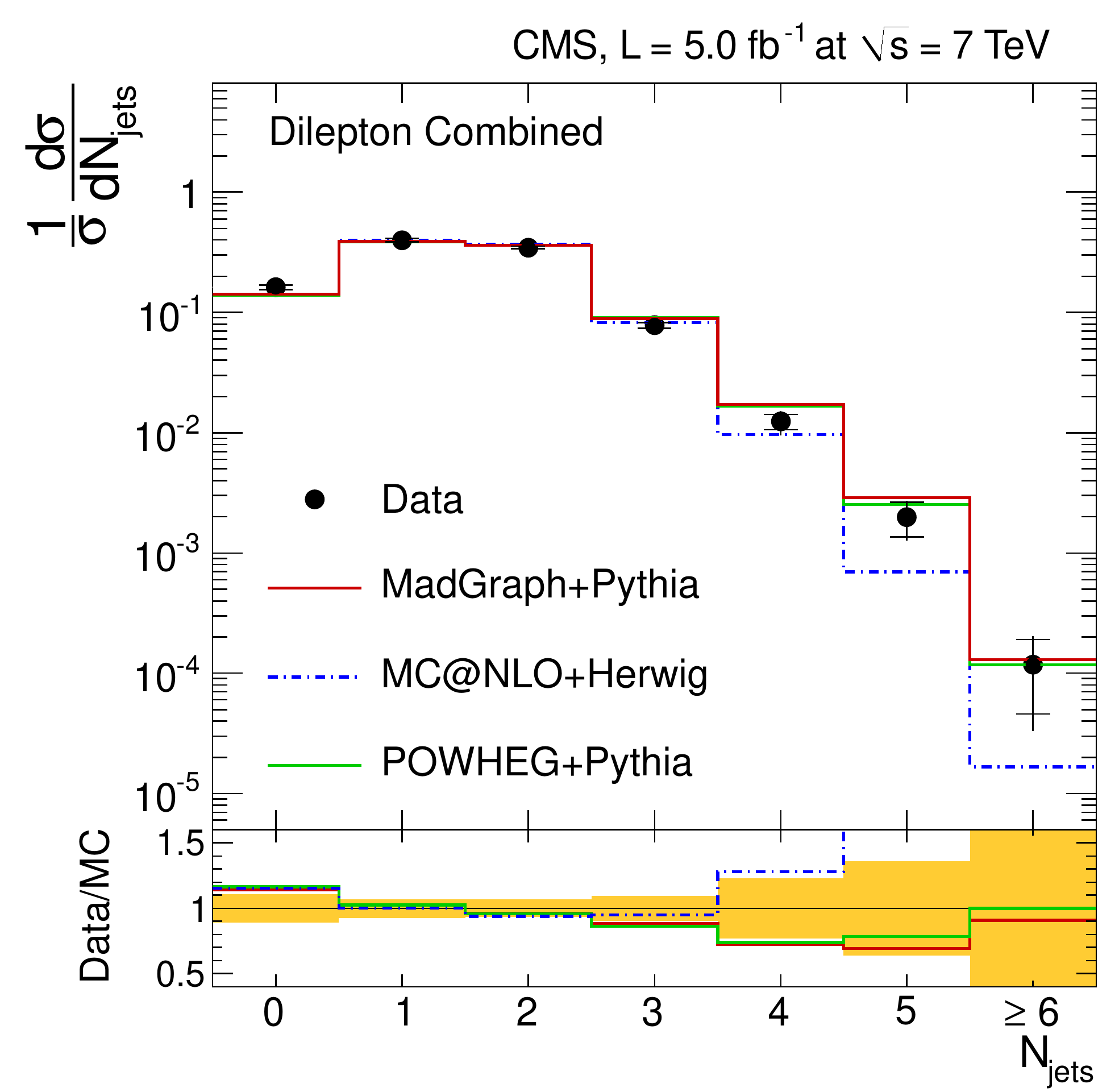}
     \includegraphics[width=0.49 \textwidth]{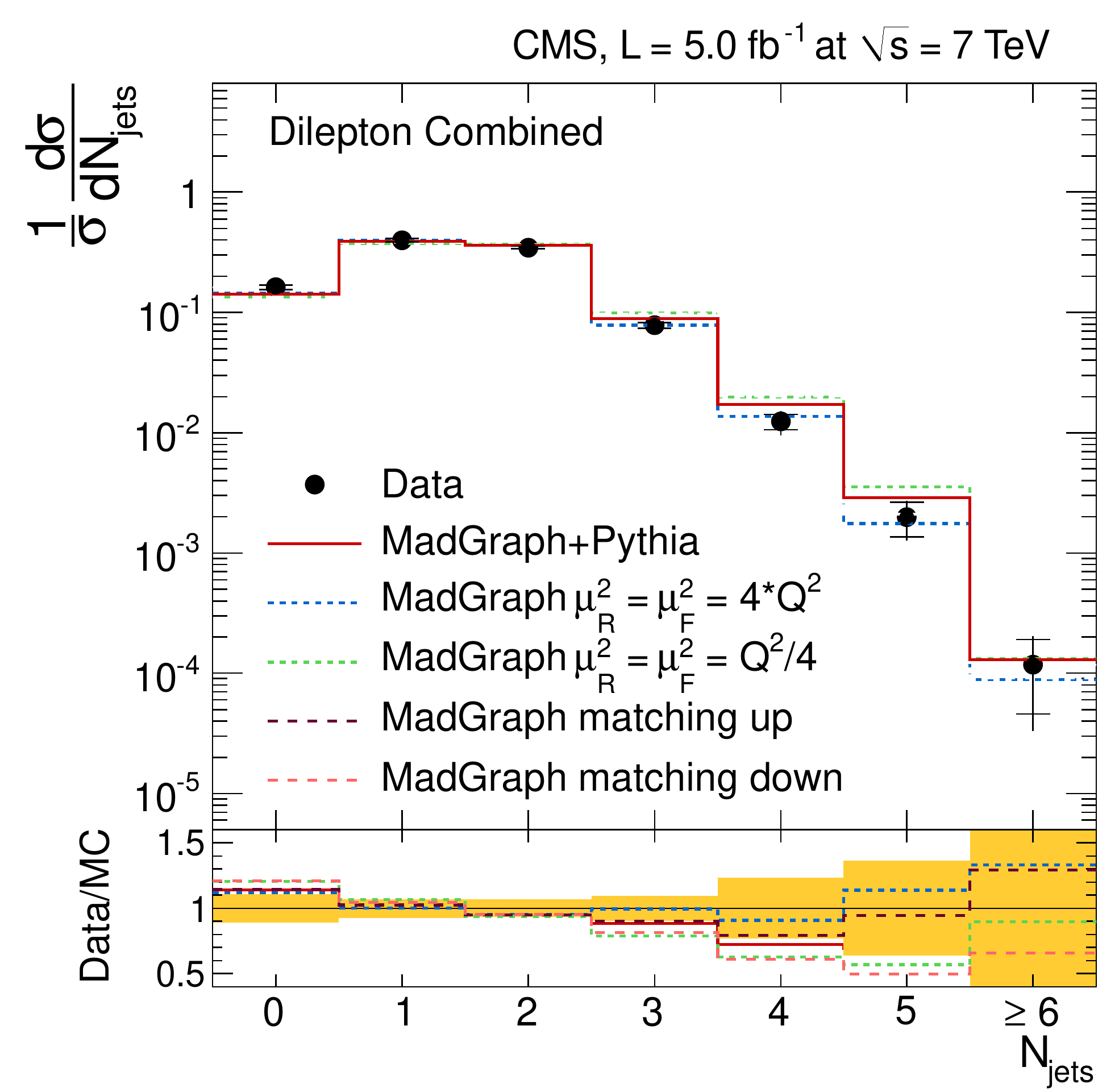}
\caption{Normalised differential \ttbar production cross section as a function of the jet multiplicity for jets with $\pt >30\GeV$ (top) and $\pt >60\GeV$ (bottom) in the dilepton channel. The measurements are compared to predictions from \MADGPYT, \POWPYT, and \MCNLOHER (left), as well as from \MADGRAPH with varied renormalisation and factorisation scales, and jet-parton matching threshold (right). The inner (outer) error bars indicate the statistical (combined statistical and systematic) uncertainty. The shaded band corresponds to the combined statistical and systematic uncertainty.}
\label{fig:xsecjet}
\end{figure*}

\begin{figure}[!ht]
\centering
\includegraphics[width=0.49\textwidth]{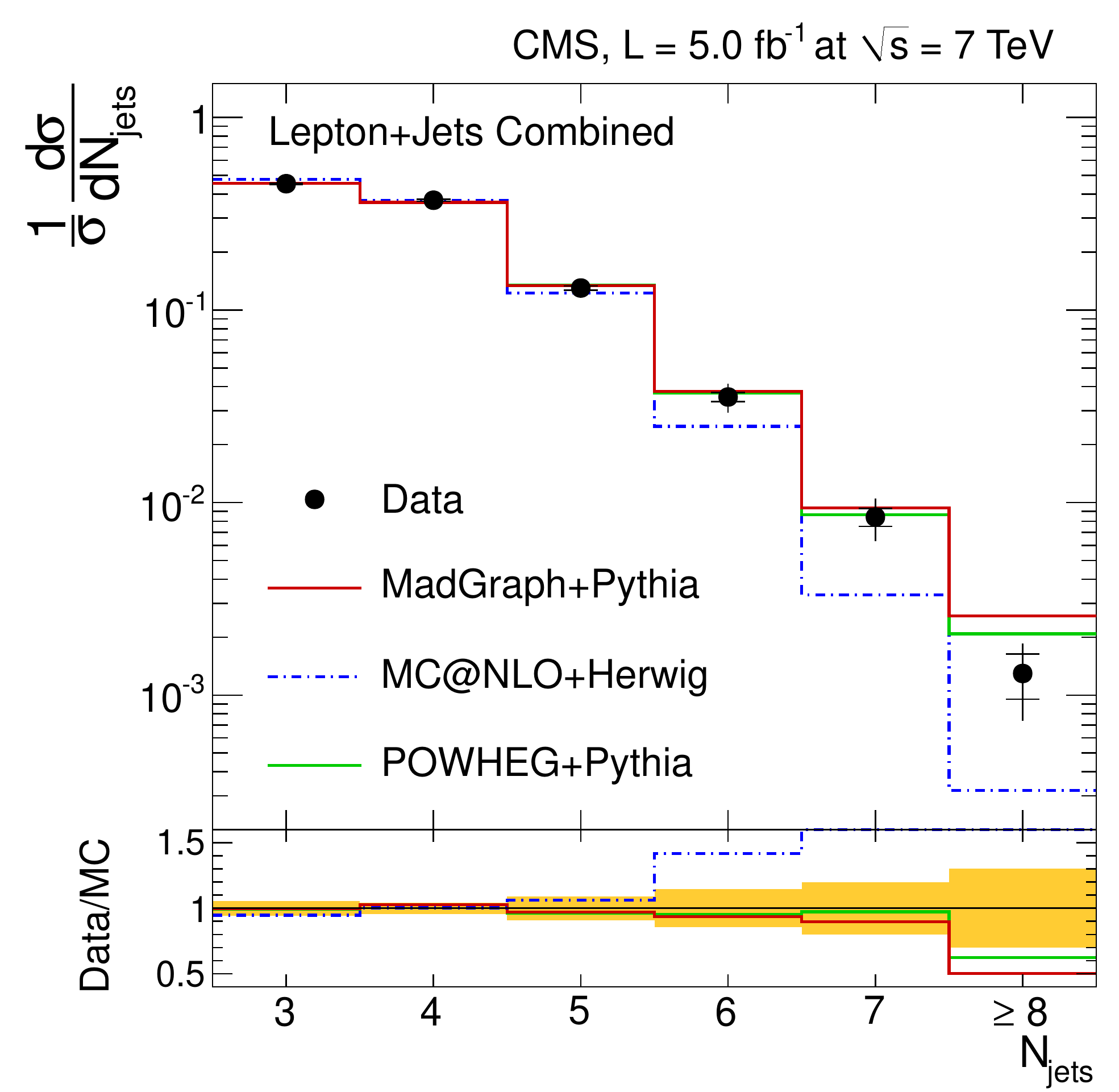}
\includegraphics[width=0.49\textwidth]{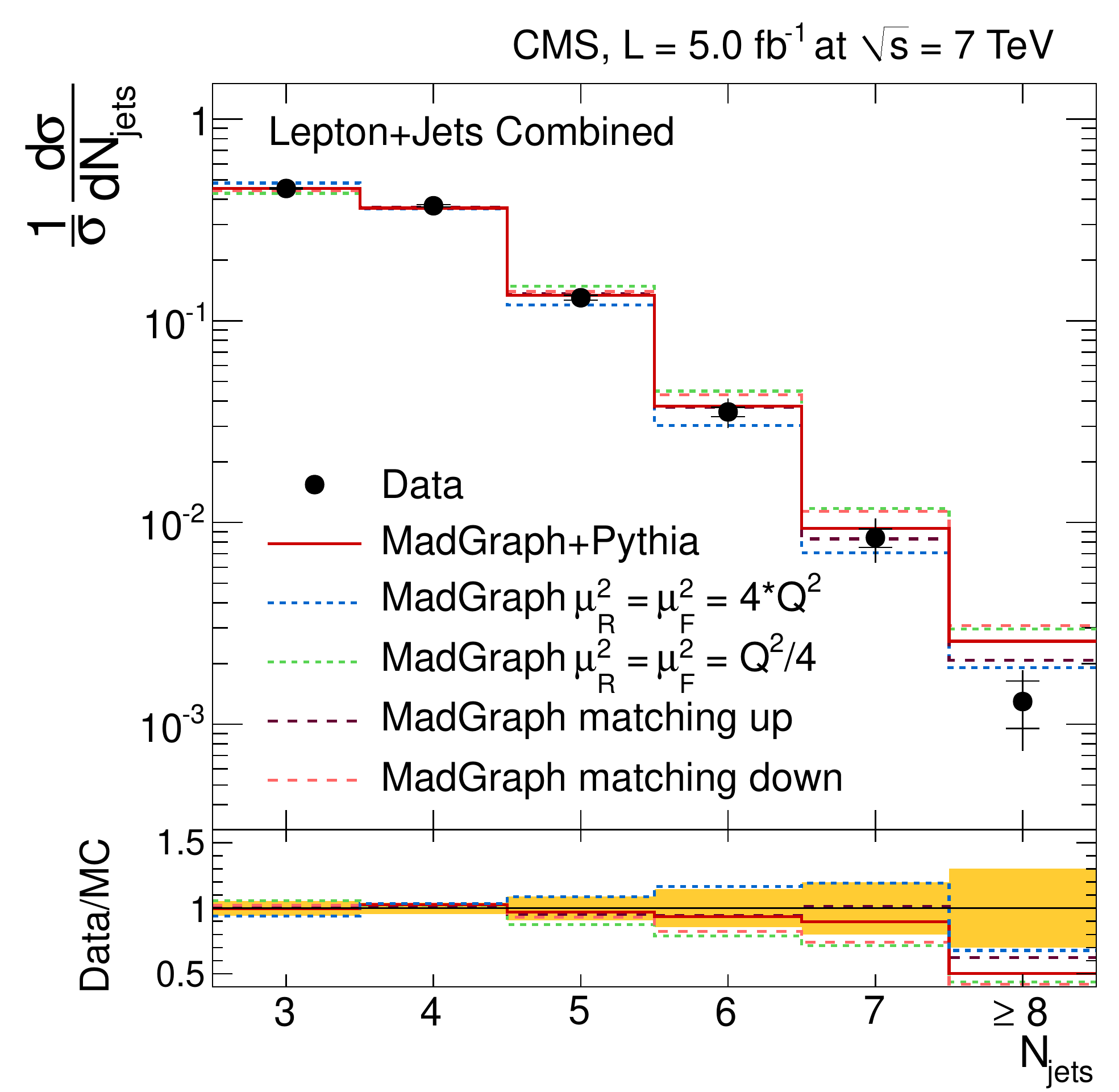}
  \caption{Normalised differential \ttbar production cross section as a function of jet multiplicity for jets with $\pt>35\GeV$ in the $\ell$+jets channel. The measurement is compared to predictions from \MADGPYT, \POWPYT, and \MCNLOHER (\cmsLeft), as well as from \MADGRAPH with varied renormalisation and factorisation scales, and jet-parton matching threshold (\cmsRight). The inner (outer) error bars indicate the statistical (combined statistical and systematic) uncertainty. The shaded band corresponds to the combined statistical and systematic uncertainty.}\label{fig:results:dXS}
\end{figure}

The normalised differential \ttbar production cross section, $1/\sigma\,\rd\sigma/\rd{}N_{\text{jets}}$, as a function of the jet multiplicity, $N_{\text{jets}}$, is shown in Tables~\ref{tab:xsecdil30} and~\ref{tab:xsecdil60}, and Fig.~\ref{fig:xsecjet} for the dilepton channel and jets with $\pt >30~(60)\GeV$. For the $\ell$+jets channel it is shown in Table~\ref{tab:results:norm_dXS} and Fig.~\ref{fig:results:dXS} for jets with $\pt >35\GeV$. In the tables, the experimental uncertainties are divided between the dominant (JES) and other (JER, b-tagging, pileup, lepton identification, isolation, and trigger efficiencies, background contribution and integrated luminosity) contributions. The model uncertainties are also divided between the dominant (renormalisation and factorisation scales, jet-parton matching threshold, and hadronisation) and other (PDF and colour reconnection) contributions.
The measurements are compared to the predictions from \MADGRAPH and \POWHEG, both interfaced with \PYTHIA, and from \MCATNLO interfaced with \HERWIG.

{\tolerance=1200
The predictions from \MADGPYT and \breakhere\POWPYT are found to provide a reasonable description of the data. In contrast, \MCNLOHER generates fewer events in bins with large jet multiplicities.
The effect of the variation of the renormalisation and factorisation scales and jet-parton matching threshold in \MADGPYT is compared with the reference \MADGPYT simulation. The choice of lower values for both these parameters seems to provide a worse description of the data for higher jet multiplicities.
\par}

\begin{table*}[htpb!]
\centering
\topcaption{Normalised differential \ttbar production cross section as a function of the jet multiplicity for jets with $\pt >30\GeV$ in the dilepton channel. The statistical, systematic, and total uncertainties are also shown. The main experimental and model systematic uncertainties are displayed: JES and the combination of renormalisation and factorisation scales, jet-parton matching threshold, and hadronisation (in the table ``$Q^2$/Match./Had.").   }
\label{tab:xsecdil30}
\begin{tabular}{c|D{.}{.}{1.3}|D{.}{.}{2.1}|D{.}{.}{2.1}D{.}{.}{1.2}|D{.}{.}{2.1}D{.}{.}{1.1}|D{.}{.}{2.1}}
\hline
\multicolumn{1}{c|}{$N_{\text{jets}}$} & \multicolumn{1}{c|}{$1/\sigma\,\rd\sigma/\rd{}N_{\text{jets}}$} & \multicolumn{1}{c|}{Stat. (\%)}  & \multicolumn{2}{c|}{ Exp. Syst. (\%)}  & \multicolumn{2}{c|}{Model Syst. (\%)} & \multicolumn{1}{c}{Total (\%)} \\
 & & & \multicolumn{1}{c}{JES}  & \multicolumn{1}{c|}{Other} & \multicolumn{1}{c}{$Q^2$/Match./Had.} & \multicolumn{1}{c|}{Other} & \\
\hline
$2$ & 0.600 & 1.2 		& 1.4 & 0.6 & 0.5 & 1.6 & 2.5\\
$3$ & 0.273 & 3.3 		& 2.3 & 2.8 & 5.4 & 1.6 & 7.2\\
$4$ & 0.096 & 5.1 		& 6.3 & 3.4 & 2.8 & 1.6 & 9.3\\
$5$ & 0.025 & 10.1 		& 7.9 & 3.0 & 17.4 & 1.9 & 24.0\\
$\geq$6 & 0.0013 & 23.8 	& 14.2 & 2.8 & 24.3 & 2.1 & 37.1\\\hline
\end{tabular}
\end{table*}

\begin{table*}[htpb!]
\centering
\topcaption{Normalised differential \ttbar production cross section as a function of the jet multiplicity for jets with $\pt >60\GeV$ in the dilepton channel. The statistical, systematic, and total uncertainties are also shown. The main experimental and model systematic uncertainties are displayed: JES and the combination of renormalisation and factorisation scales, jet-parton matching threshold, and hadronisation (in the table ``$Q^2$/Match./Had.").}
\label{tab:xsecdil60}
\begin{tabular}{c|D{.}{.}{1.5}|D{.}{.}{2.1}|D{.}{.}{1.1}D{.}{.}{1.2}|D{.}{.}{2.1}D{.}{.}{1.1}|D{.}{.}{2.1}}
\hline
\multicolumn{1}{c|}{$N_{\text{jets}}$} & \multicolumn{1}{c|}{$1/\sigma\,\rd\sigma/\rd{}N_{\text{jets}}$} & \multicolumn{1}{c|}{Stat. (\%)}  & \multicolumn{2}{c|}{ Exp. Syst. (\%)}  & \multicolumn{2}{c|}{Model Syst. (\%)} & \multicolumn{1}{c}{Total (\%)} \\
 & & & \multicolumn{1}{c}{JES}  & \multicolumn{1}{c|}{Other} & \multicolumn{1}{c}{$Q^2$/Match./Had.} & \multicolumn{1}{c|}{Other} & \\
\hline
$0$ &  0.158 & 3.4 		& 7.0 & 5.7 & 2.7 & 1.6 & 10.1\\
$1$ &  0.397 & 4.0 		& 4.9 & 2.0 & 3.3 & 1.9 & 7.6\\
$2$ &  0.350 & 2.6 		& 3.2 & 3.3 & 3.5 & 1.7 & 6.6\\
$3$ &  0.079 & 5.2 		& 3.4 & 3.0 & 5.8 & 1.6 & 9.2\\
$4$ &  0.0127 & 13.9 		&  5.4 & 3.5 & 15.8 & 1.7 & 22.1\\
$5$ &  0.0020 & 30.9 		& 4.8 & 3.6 & 15.5 & 1.6 & 35.1\\
$\geq$6 & 0.00012 & 57.1 	& 4.7 & 16.7 & 38.7 & 2.9 & 69.4\\\hline
\end{tabular}
\end{table*}

\begin{table*}[htpb!]
\centering
\topcaption{Normalised differential \ttbar production cross section as a function of the jet multiplicity for jets with $\pt >35\GeV$ in the $\ell$+jets channel. The statistical, systematic, and total uncertainties are also shown. The main experimental and model systematic uncertainties are displayed: JES and the combination of renormalisation and factorisation scales, jet-parton matching threshold, and hadronisation (in the table ``$Q^2$/Match./Had."). }\label{tab:results:norm_dXS}
\begin{tabular}{c|D{.}{.}{1.5}|D{.}{.}{2.1}|D{.}{.}{2.1}D{.}{.}{1.2}|D{.}{.}{2.1}D{.}{.}{1.1}|D{.}{.}{2.1}}
\hline
\multicolumn{1}{c|}{$N_{\text{jets}}$} & \multicolumn{1}{c|}{$1/\sigma\,\rd\sigma/\rd{}N_{\text{jets}}$} & \multicolumn{1}{c|}{Stat. (\%)}  & \multicolumn{2}{c|}{ Exp. Syst. (\%)}  & \multicolumn{2}{c|}{Model Syst. (\%)} & \multicolumn{1}{c}{Total (\%)} \\
 & & & \multicolumn{1}{c}{JES}  & \multicolumn{1}{c|}{Other} & \multicolumn{1}{c}{$Q^2$/Match./Had.} & \multicolumn{1}{c|}{Other} & \\
\hline
$3$ & 0.453 & 0.9 		& 3.8 	& 2.2	& 3.8 	& 1.3	& 6.1\\
$4$ & 0.372 & 1.2 		& 1.8 	& 1.8	& 3.2 	& 1.4	& 4.5\\
$5$ & 0.130 & 2.7 		& 5.6 	& 2.0	& 7.5 	& 1.8	& 10.2\\
$6$ & 0.0353 & 5.3 		& 6.7 	& 2.4	& 14.2	& 2.5	& 17.0\\
$7$ & 0.00841 & 10.5 		& 10.7	& 3.3	& 19.1	& 4.3	& 24.9\\
$\geq$8 & 0.00130 & 26.4 	& 17.7	& 5.1	& 28.6	& 3.4	& 43.2\\\hline
\end{tabular}
\end{table*}

\section{Normalised differential cross section as a function of the additional jet multiplicity}
\label{sec:addJets}

The normalised differential \ttbar production cross section is also determined as a function of the number of additional jets accompanying the \ttbar decays in the $\ell$+jets channel. This measurement provides added value to the one presented in Sect.~\ref{sec:results} by distinguishing jets from the \ttbar decay products and jets coming from additional QCD radiation. This is particularly interesting in final states with many jets.

\begin{figure}[htb!]
\centering
		\includegraphics[width=0.49\textwidth]{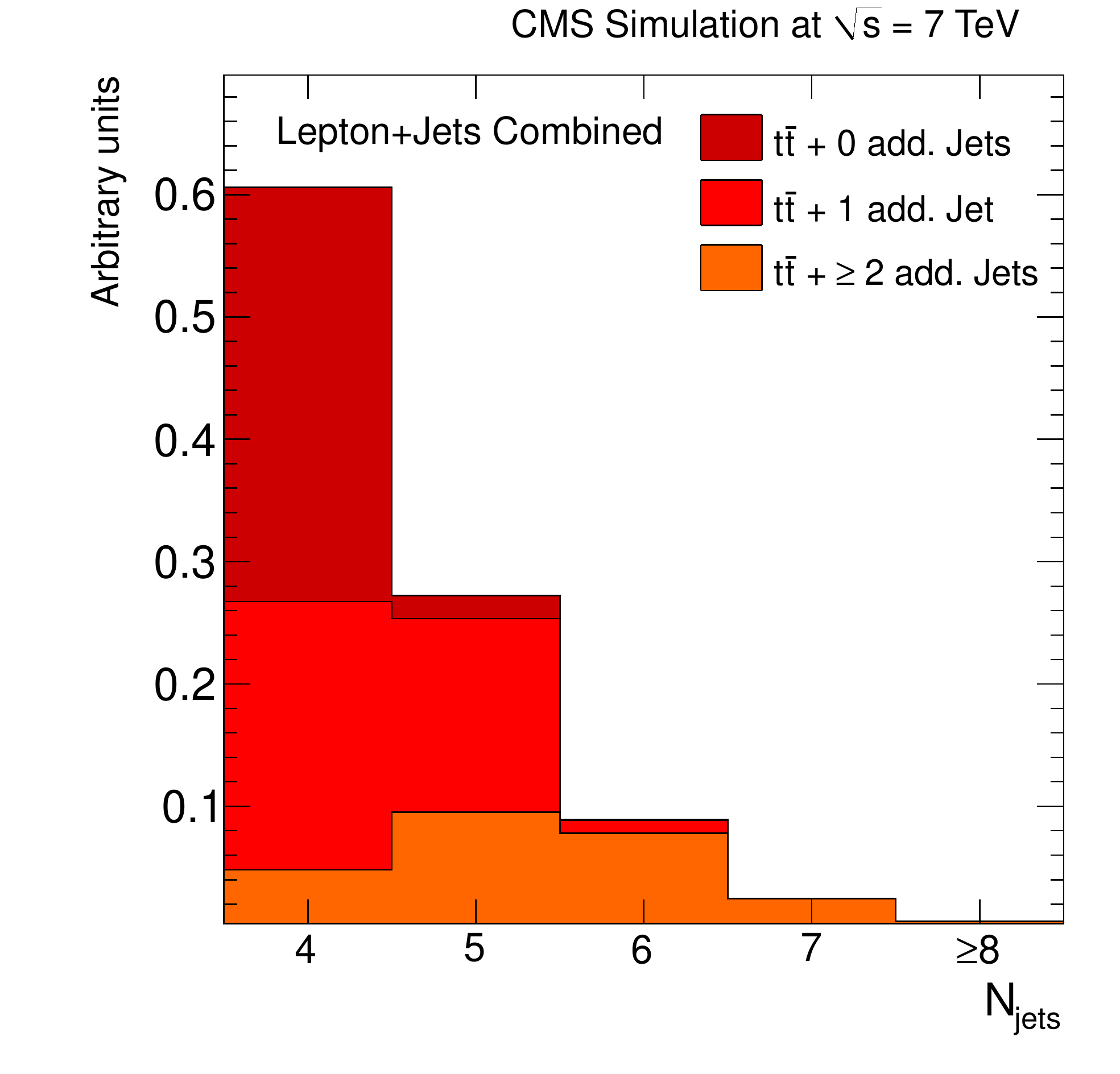}
		\caption{Jet multiplicity distribution in simulated \ttbar events in the $\ell$+jets channel. The splitting into three categories, defined by the compatibility of the selected particle level jets with the \ttbar decay partons is also shown (cf. Sect.~7).\label{fig:split_njet}}
\end{figure}

For this measurement, the event selection follows the prescription discussed in
Sect.~\ref{sec:selection}, and requires at least four jets (in order to perform a full event reconstruction later) with $\pt >30\GeV$ and $\abs{\eta}<2.4$. The \pt requirement is lowered to gain more data and reduce the statistical uncertainty. The particle-level jets, defined as described in Sect.~\ref{sec:results} but with $\pt >30\GeV$, are counted as additional jets if their distance to the \ttbar decay products is $\Delta R>0.5$. We consider the following objects as \ttbar decay products: two b quarks, two light quarks from the hadronically decaying $W$ boson, and the lepton from the leptonically decaying $W$ boson; the neutrino is not included. The simulated \ttbar events are classified into three categories according to the number of additional jets (0, 1, and $\geq$2) selected according to this definition.
Figure~\ref{fig:split_njet} illustrates the contributions of \ttbar events with 0, 1, and $\geq$2 additional jets to the number of reconstructed jets in the simulation.

A full event reconstruction of the \ttbar system is performed in order to create a variable sensitive to additional jets, taking into account all possible jet permutations. The most likely permutation is determined using a $\chi^2$ minimisation, where the $\chi^2$ is given by:
\ifthenelse{\boolean{cms@external}}{
\begin{multline*}
\chi^2 = \left(\frac{m^{\text{rec}}_{\PW^{\text{had}}} - m^{\text{true}}_{\PW^{\text{had}}}}{\sigma_{\PW^{\text{had}}}}\right)^2
\\
+ \left(\frac{m^{\text{rec}}_{\cPqt^{\text{had}}} - m^{\text{true}}_{\cPqt^{\text{had}}}}{\sigma_{\cPqt^{\text{had}}}}\right)^2
+ \left(\frac{m^{\text{rec}}_{\cPqt^{\text{lep}}} - m^{\text{true}}_{\cPqt^{\text{lep}}}}{\sigma_{\cPqt^{\text{lep}}}}\right)^2,
\end{multline*}
}{
\begin{equation*}
\chi^2 = \left(\frac{m^{\text{rec}}_{\PW^{\text{had}}} - m^{\text{true}}_{\PW^{\text{had}}}}{\sigma_{\PW^{\text{had}}}}\right)^2 + \left(\frac{m^{\text{rec}}_{\cPqt^{\text{had}}} - m^{\text{true}}_{\cPqt^{\text{had}}}}{\sigma_{\cPqt^{\text{had}}}}\right)^2  + \left(\frac{m^{\text{rec}}_{\cPqt^{\text{lep}}} - m^{\text{true}}_{\cPqt^{\text{lep}}}}{\sigma_{\cPqt^{\text{lep}}}}\right)^2,
\end{equation*}
}
where $m_{\cPqt^{\text{had}}}^{\text{rec}}$ and $m_{\cPqt^{\text{lep}}}^{\text{rec}}$ are the reconstructed invariant masses of the hadronically and the leptonically decaying top quark, respectively, and $m_{\PW^\text{had}}$ is the reconstructed invariant mass of the W boson from the hadronic top-quark decay. The parameters $m^{\text{true}}$ and $\sigma_{\cPqt^{\text{had}}}$, $\sigma_{\cPqt^{\text{lep}}}$, and $\sigma_{\PW^{\text{had}}}$ are the mean value and standard deviations of the reconstructed mass distributions in the \ttbar simulation. In each event, all jet permutations in which only b-tagged jets are assigned to b quarks are considered. The permutation with the smallest $\chi^2$ value is chosen as the best hypothesis. For events containing the same number of reconstructed jets ($N_\text{jets}$) the variable $\sqrt{\chi^2}$ provides good discrimination between events classified as \ttbar + 0, 1, and $\geq$2 additional jets. The discrimination power is due to the sensitivity of the event reconstruction to the relation between $N_\text{jets}$ and the number of additional jets $N_\text{add. jets}$. The best event reconstruction, thus providing a smaller $\sqrt{\chi^2}$, is achieved if the observation is close to $N_\text{jets} = 4 + N_\text{add. jets}$, where four is the expected number of jets from the \ttbar decay partons. For instance, a \ttbar + 1 additional jet event with $N_\text{jets}=4$ is likely to get a large $\sqrt{\chi^2}$ value because one of the four jets from the \ttbar decay partons is missing for a correct event reconstruction.

\begin{figure*}[htb!]
	\centering
	\includegraphics[width=\cmsFigWidth]{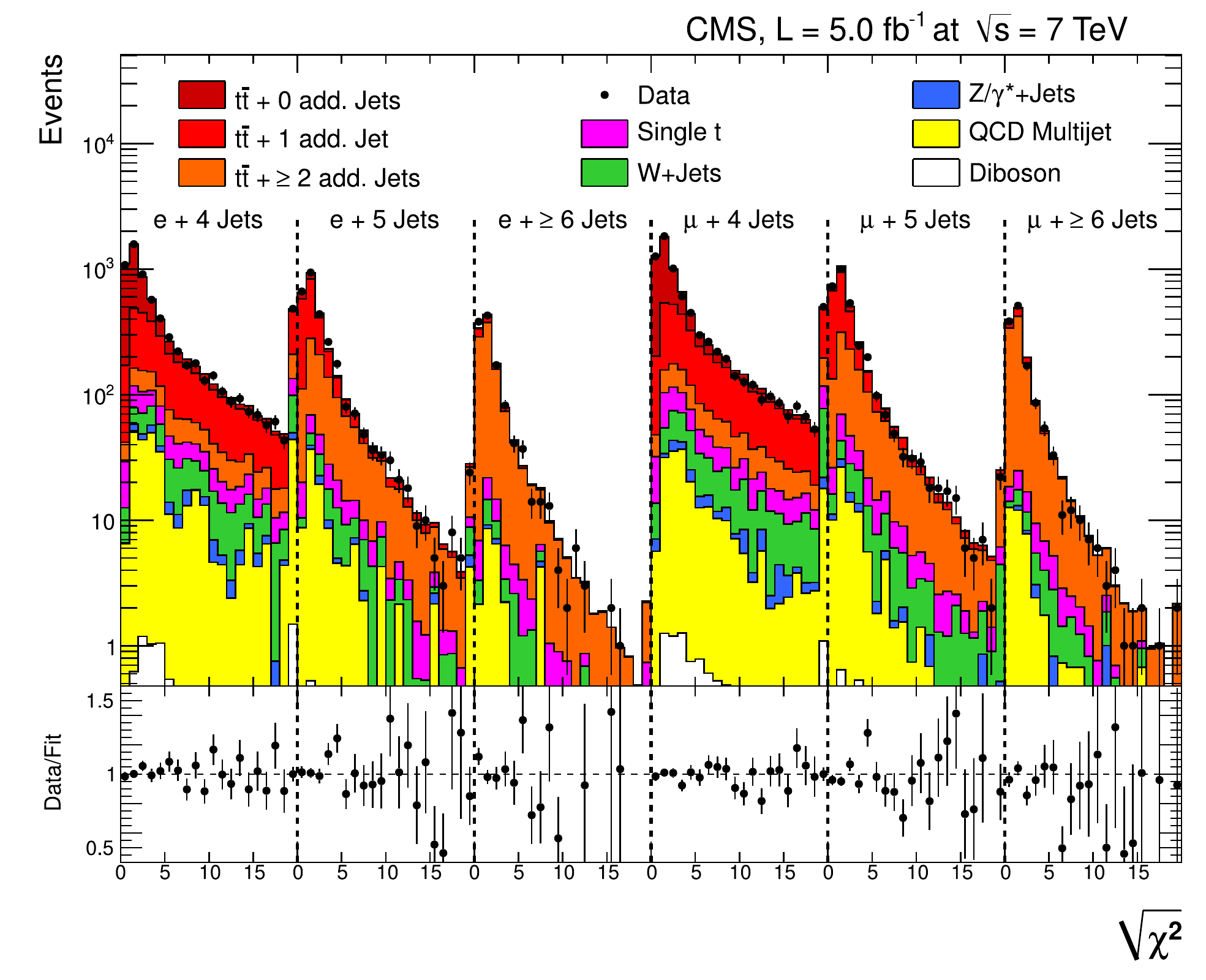}
	\caption{Result of the simultaneous template fit to the $\sqrt{\chi^2}$ distribution in the $\ell$+jets channel. All templates are scaled to the resulting fit parameters. \label{fig:unfolded_meas}}
\end{figure*}

The measurement of the fractions of \ttbar events with 0, 1, and $\geq$2 additional jets is performed using a binned maximum-likelihood fit of the $\sqrt{\chi^2}$ templates to data, simultaneously in both $\ell$+jets channels. The normalisations of the signal templates (\ttbar + 0, 1, and $\geq$2 additional jets) are free parameters in the fit. For the normalisations of the background processes, Gaussian constraints corresponding to the uncertainties of the background predictions are applied. It has been verified that the use of log-normal constraints give similar results. The result of the fit is shown in Fig.~\ref{fig:unfolded_meas}. The QCD multijet and W+jets templates are estimated using the data-based methods described in Sect.~\ref{sec:selection}.

\begin{figure}[htb!]
	\centering
	\includegraphics[width=0.49 \textwidth]{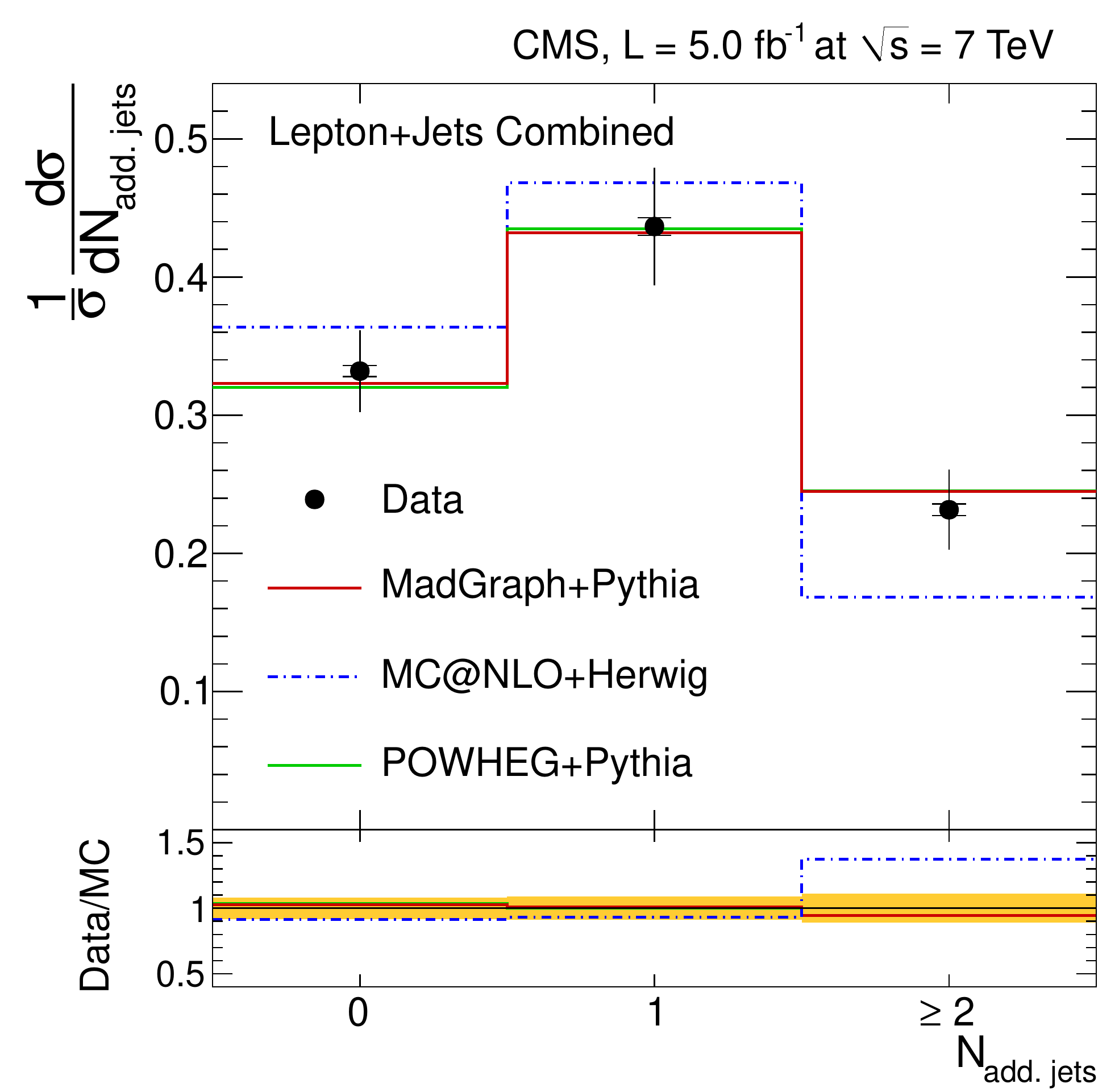}
    \includegraphics[width=0.49 \textwidth]{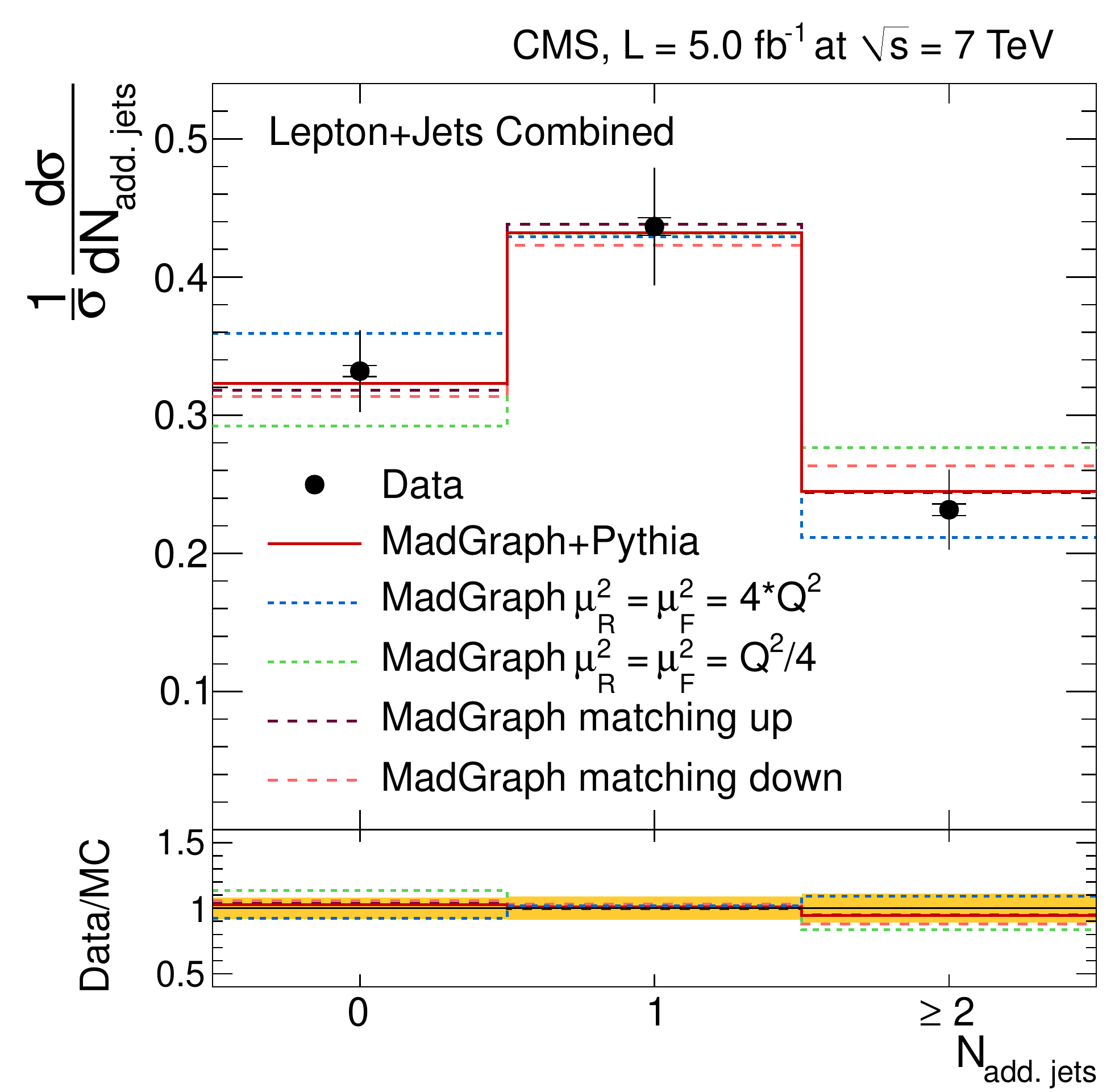}
	\caption{Normalised differential \ttbar production cross section as a function of the number of additional jets in the $\ell$+jets channel. The measurement is compared to predictions from \MADGPYT, \POWPYT, and \MCNLOHER (\cmsLeft), as well as from \MADGRAPH with varied renormalisation and factorisation scales, and jet-parton matching threshold (\cmsRight). The inner (outer) error bars indicate the statistical (combined statistical and systematic) uncertainty. The shaded band corresponds to the combined statistical and systematic uncertainty\label{fig:unfolded_meas_2}}
\end{figure}

{\tolerance=800
The normalisations for the three signal templates are applied to the predicted differential cross section in the visible phase space, calculated using the simulated \ttbar sample from  \textsc{MadGraph+pythia}. This phase space is defined as in Sect.~\ref{sec:results} with the requirement of four particle level jets with $\pt>30\GeV$. This provides the differential cross section as a function of the number of additional jets, which is finally normalised to the total cross section measured in the same phase space. The results are shown in Fig.~\ref{fig:unfolded_meas_2} and summarised in Table~\ref{tab:unfold_results}.

For each \ttbar + additional jet template used in the maximum-likelihood fit, a full correlation is assumed between the rate of events that fulfill the particle-level selection and the rate of events that do not. Therefore, a single template is used for both parts.
\par}

Including an additional template made from events that are not inside the visible phase space leads to fit results that are compatible within the estimated uncertainties. To check the model dependency, the fit is repeated using simulated data from \MCNLOHER and \POWPYT instead of \MADGPYT. The results are stable within the uncertainties.

\begin{table*}[htpb!]
\centering
\topcaption{Normalised differential \ttbar production cross section as a function of the jet multiplicity for jets with $\pt >30\GeV$ in the dilepton channel. The statistical, systematic, and total uncertainties are also shown. The main experimental and model systematic uncertainties are displayed: JES and the combination of renormalisation and factorisation scales, jet-parton matching threshold, and hadronisation (in the table ``$Q^2$/Match./Had.")
\label {tab:unfold_results}}
\begin{footnotesize}
\begin{tabular}{c|D{.}{.}{1.5}|D{.}{.}{2.1}|D{.}{.}{2.1}D{.}{.}{1.2}|D{.}{.}{2.1}D{.}{.}{1.1}|D{.}{.}{2.1}}
\hline
\multicolumn{1}{c|}{$N_{\text{jets}}$} & \multicolumn{1}{c|}{$1/\sigma\,\rd\sigma/\rd{}N_{\text{add. jets}}$} & \multicolumn{1}{c|}{Stat. (\%)}  & \multicolumn{2}{c|}{ Exp. Syst. (\%)}  & \multicolumn{2}{c|}{Model Syst. (\%)} & \multicolumn{1}{c}{Total (\%)} \\
 & & & \multicolumn{1}{c}{JES}  & \multicolumn{1}{c|}{Other} & \multicolumn{1}{c}{$Q^2$/Match./Had.} & \multicolumn{1}{c|}{Other} & \\
\hline
\ttbar+ 0 add. Jets			 &  0.332  &1.2	&4.2 	&1.4 	&7.5 	&1.6	 &9.0 \\
\ttbar+ 1 add. Jet 			&  0.436  &1.5 	&0.9 	&1.0 	&9.5 	&1.3	 &9.8 \\
\ttbar+ $\geq$2 add. Jets 		&  0.232   &1.8	&7.2 	&1.5 	&9.6 	&2.6 	 &12.5 \\
\hline
\end{tabular}
\end{footnotesize}
\end{table*}

The sources of systematic uncertainties are the same as those discussed in Sect.~\ref{sec:syst}, except for the background normalisations, which are constrained in the fit. Their effect is propagated to the fit uncertainty, which is quoted as the statistical uncertainty. The impact of the systematic uncertainties on the extracted fractions of \ttbar + 0, 1, and $\geq$2 additional jets is evaluated using pseudo-experiments.
The most important contributions to the systematic uncertainties originate from JES (up to 7\%) and modelling uncertainties: hadronisation (up to 6\%), jet-parton matching threshold (up to 5\%), and renormalisation and factorisation scales (up to 4\%).

The \MCNLOHER prediction produces fewer events with $\geq$2 additional jets than data, which are well described by \MADGPYT and \POWPYT. The prediction from \MADGPYT with lower renormalisation and factorisation scales provides a worse description of the data. These observations are in agreement with those presented in Sect.~\ref{sec:results}.

\section{Additional jet gap fraction}
\label{sec:gap}

An alternative way to investigate the jet activity arising from quark and gluon radiation produced in association with the \ttbar system is to determine the fraction of events that do not contain additional jets above a given threshold. This measurement is performed using events in the dilepton decay channel after fulfilling the event reconstruction and selection requirements discussed in Sect.~\ref{sec:selection}. The additional jets are defined as those not assigned to the \ttbar system by the kinematic reconstruction described in Sect.~\ref{subsec:selection}.

\begin{figure*}[htb!]
    \centering
        \includegraphics[width=0.49 \textwidth]{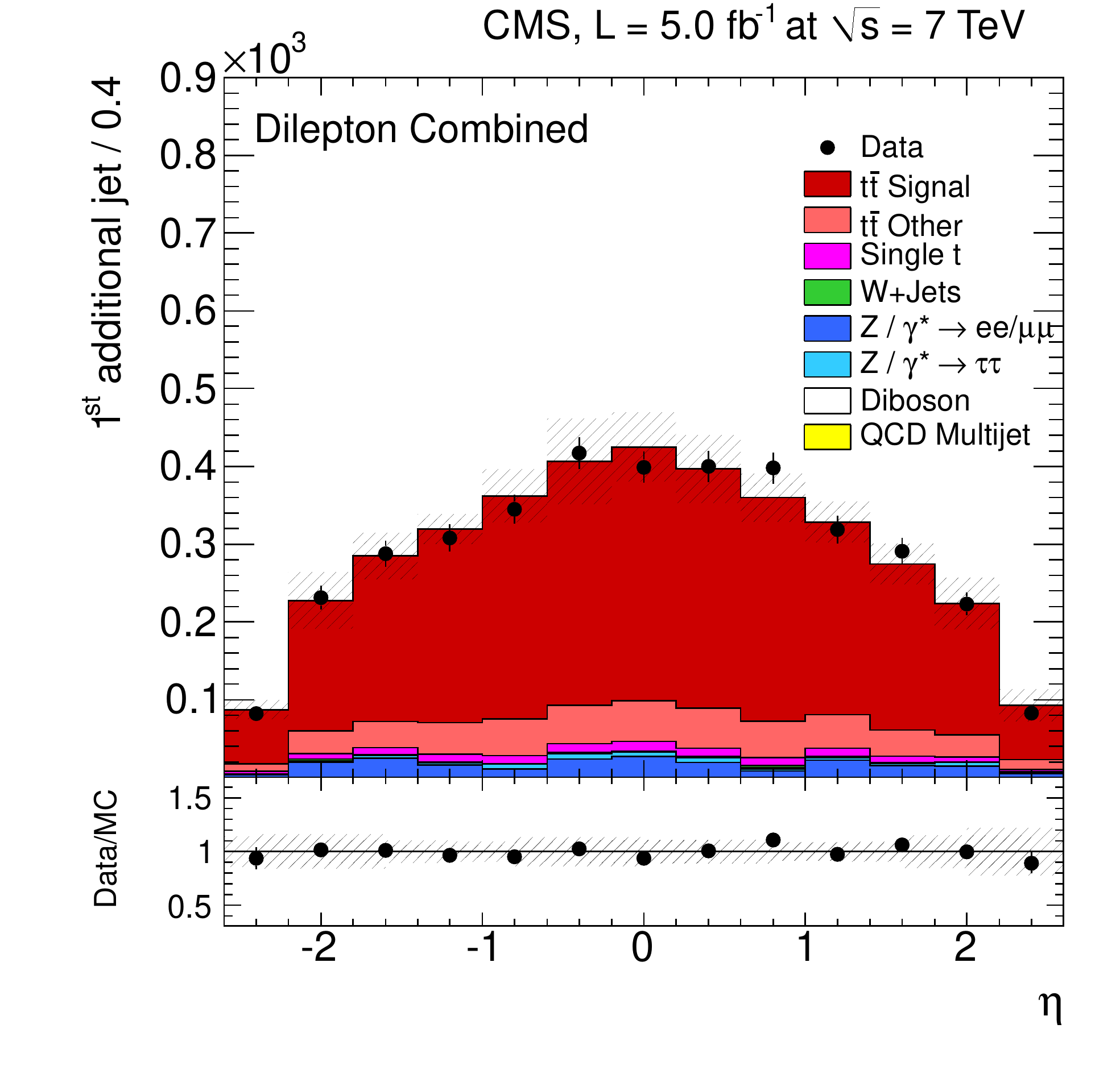}
        \includegraphics[width=0.49 \textwidth]{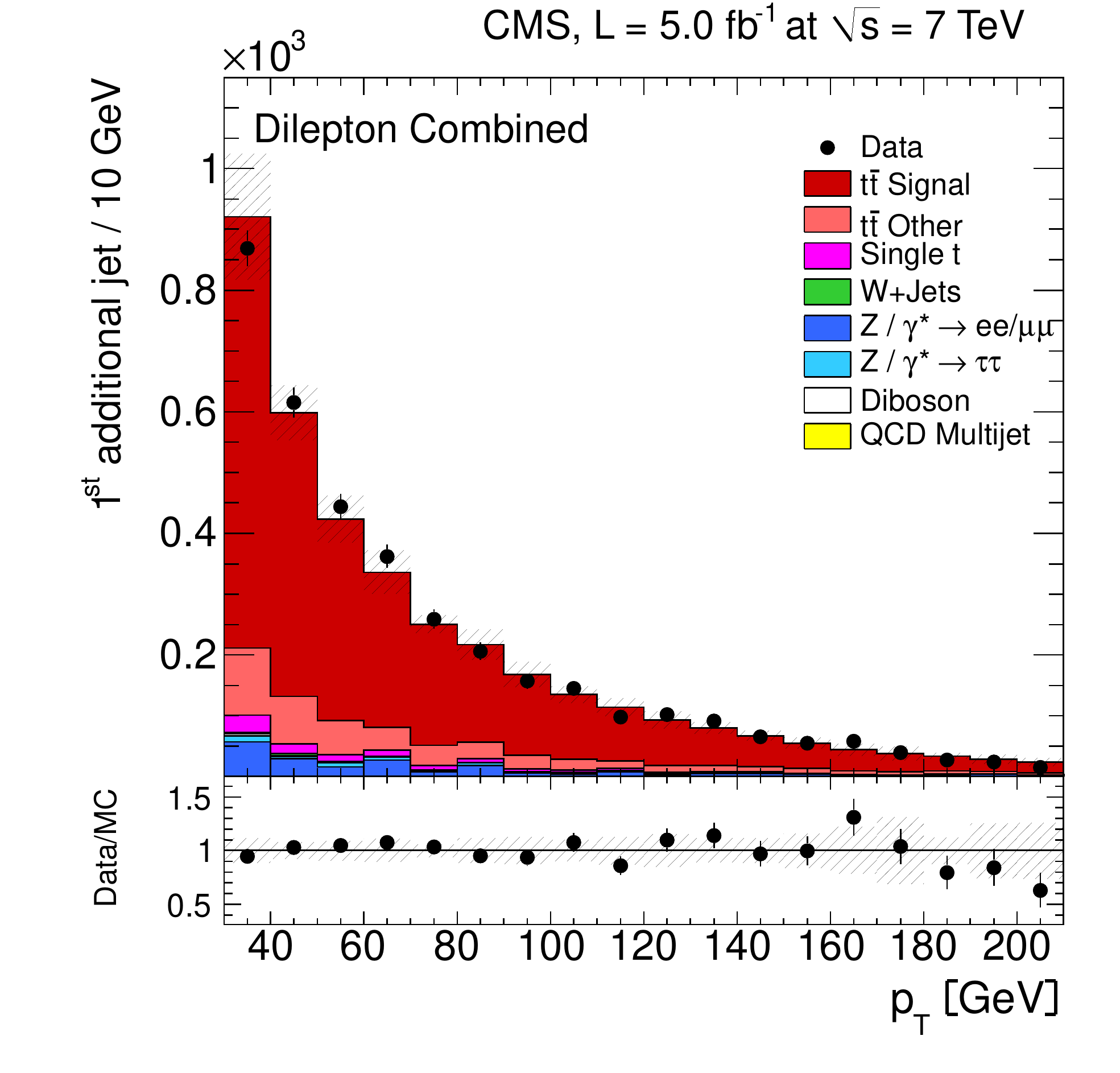}\\
        \includegraphics[width=0.49 \textwidth]{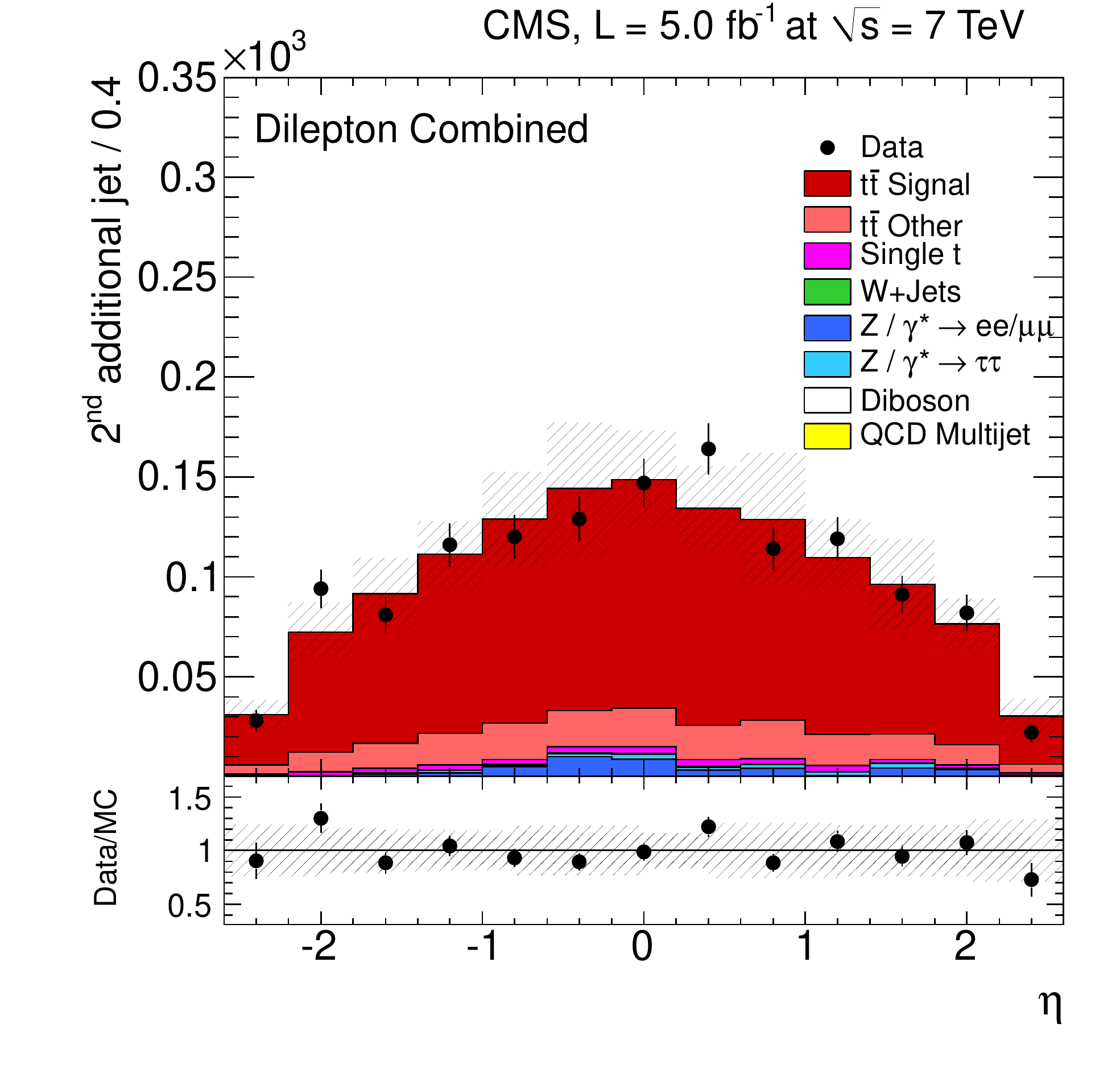}
        \includegraphics[width=0.49 \textwidth]{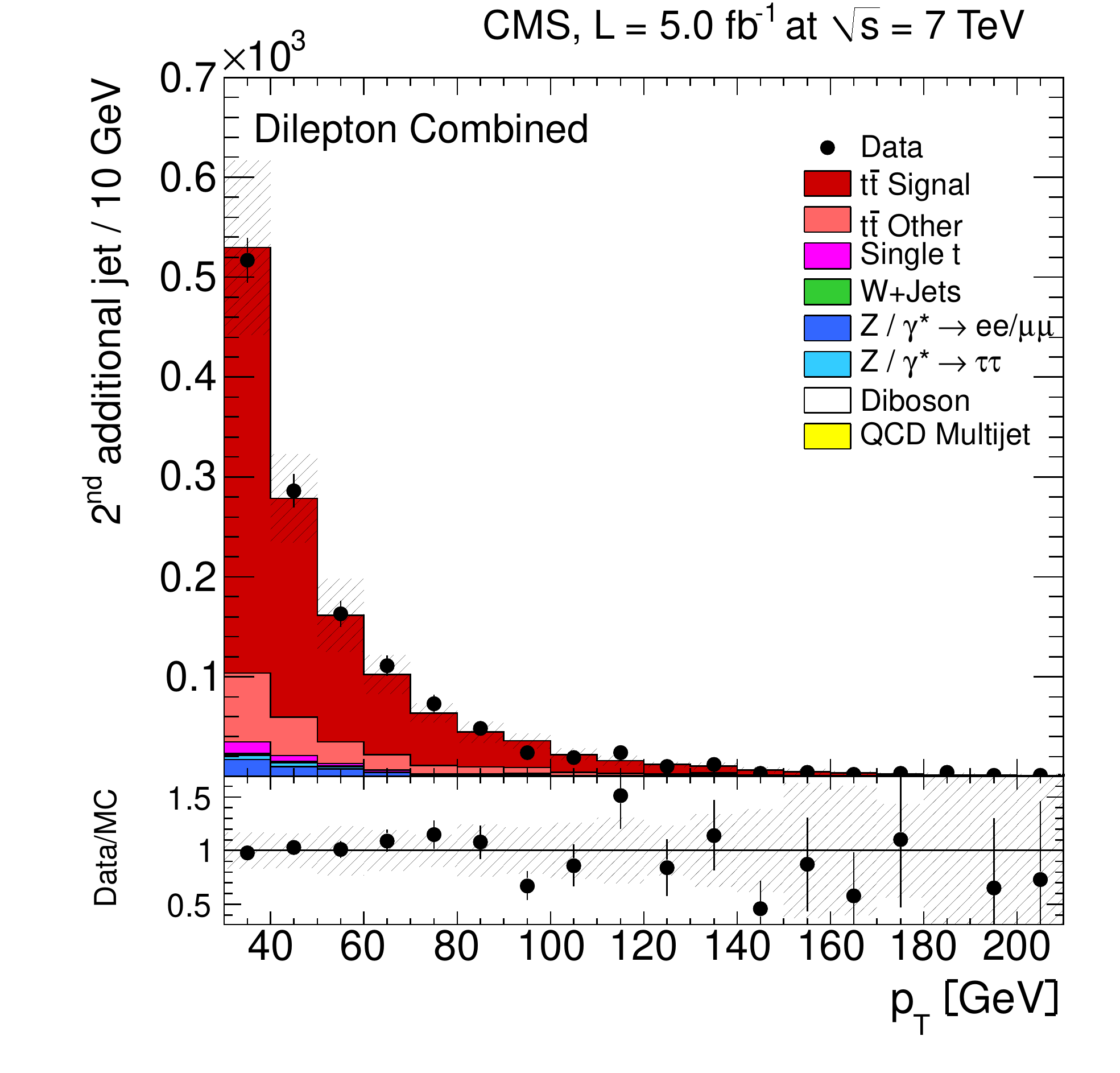}
  \caption{Distribution of the $\eta$ (left) and the \pt (right) of the first (top) and second (bottom) leading additional reconstructed jets compared to signal and background simulated samples. The error bars on the data points indicate the statistical uncertainty. The hatched band represents the combined effect of all sources of systematic uncertainty.}  \label{fig:leadjet}
   \end{figure*}

A threshold observable, referred to as gap fraction~\cite{bib:atlas2}, is defined as:
\begin{equation}
f(\pt)=\frac{N(\pt)}{N_{\text{total}}},
\end{equation}
where $N_{\text{total}}$ is the number of selected events and $N(\pt)$ is the number of events that do not contain additional jets above a \pt threshold in the whole pseudorapidity range used in the analysis ($\abs{\eta}<2.4$). The pseudorapidity and \pt distributions of the first and second leading (in \pt) additional reconstructed jets are presented in Fig.~\ref{fig:leadjet}. The distributions show good agreement between data and the simulation.

The veto can be extended beyond the additional leading jet criteria by defining the gap fraction as
\begin{equation}
f(\HT)=\frac{N(\HT)}{N_{\text{total}}},
\end{equation}
where $N(\HT)$ is the number of events in which $\HT$, the scalar sum of the \pt of the additional jets (with $\pt >30\GeV$), is less than a certain threshold.

For each value of \pt and $\HT$ thresholds, the gap fraction is evaluated at particle level in the visible phase space defined in Sect.~\ref{sec:results}.  The additional jets at particle level are defined as all jets within the kinematic acceptance not including the two highest-\pt b jets containing the decay products of different b hadrons. They are required to fulfill the condition that they are not within a cone of $\Delta R=0.4$ from any of the two isolated leptons, as described in Sect.~\ref{sec:results}.

Given the large purity of the selected events for any value of \pt and $\HT$, a correction for detector effects is applied following a simpler approach than the unfolding method used in Sect.~\ref{sec:results}. Here, the ratio of the particle-level to the simulated gap fraction distributions, obtained with the \ttbar sample from \MADGRAPH, provides the correction which is applied to the data.

\begin{figure*}[htb!]
  \centering
      \includegraphics[width=0.49 \textwidth]{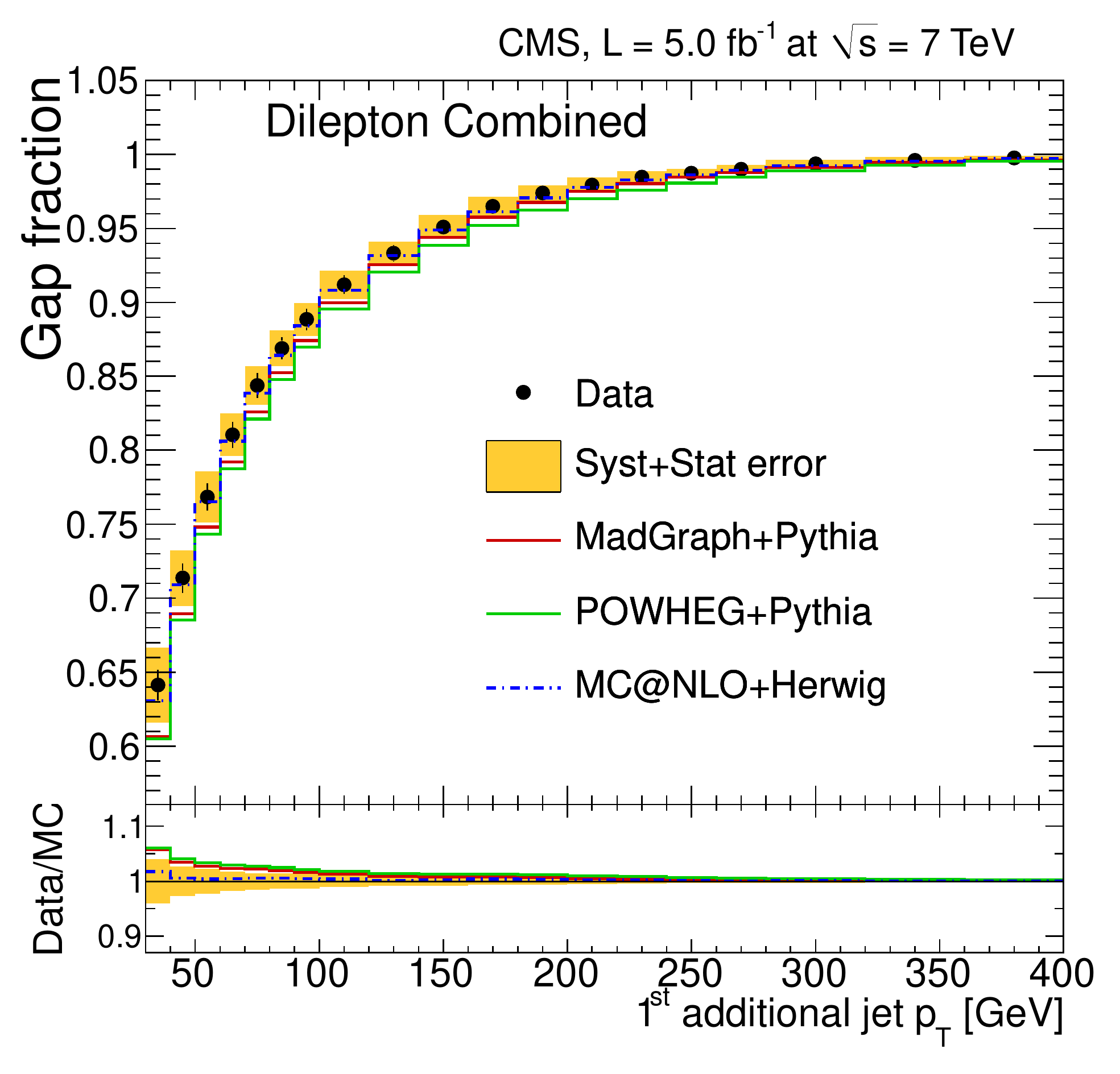}
      \includegraphics[width=0.49 \textwidth]{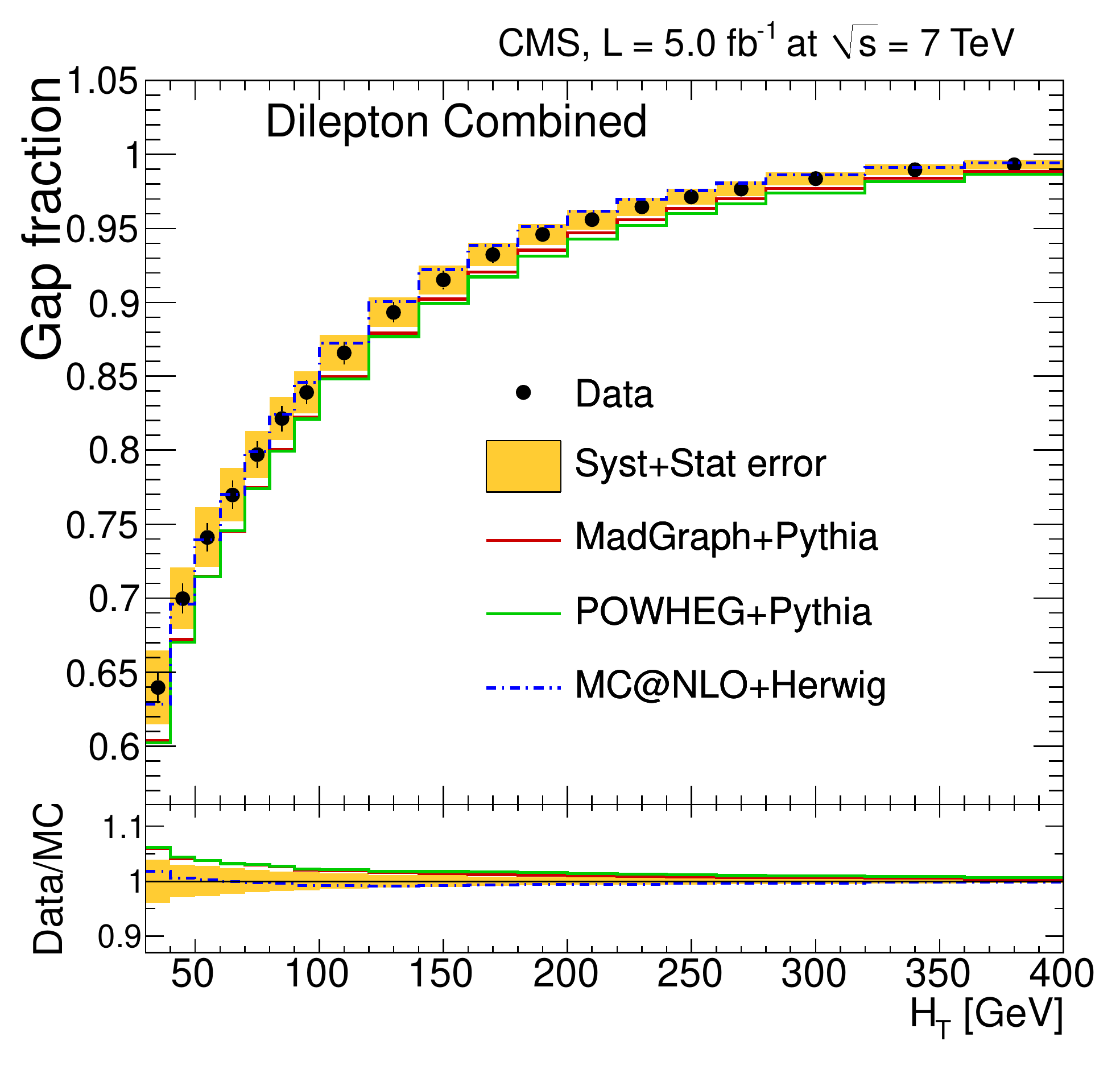}\\
      \includegraphics[width=0.49 \textwidth]{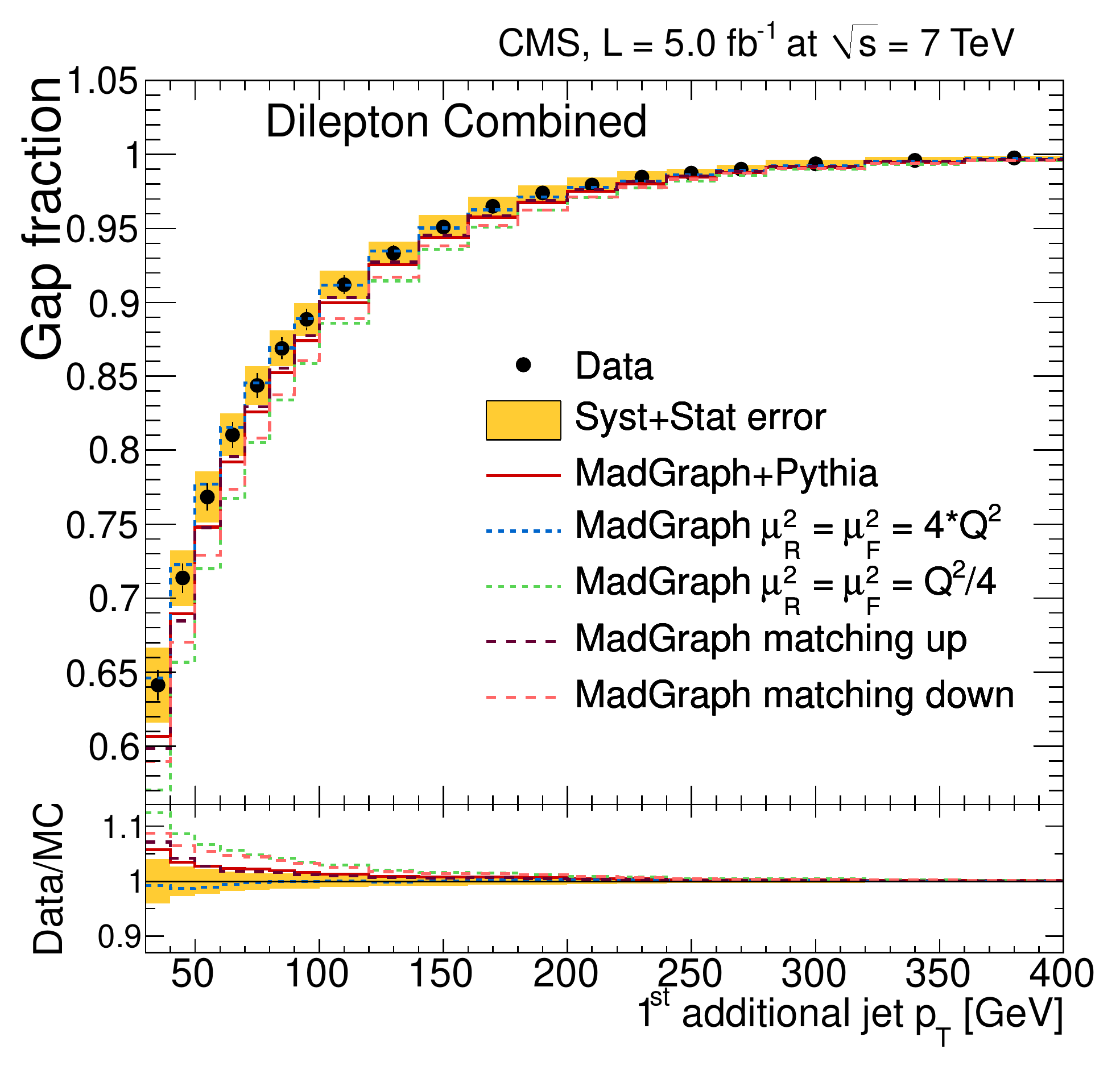}
      \includegraphics[width=0.49 \textwidth]{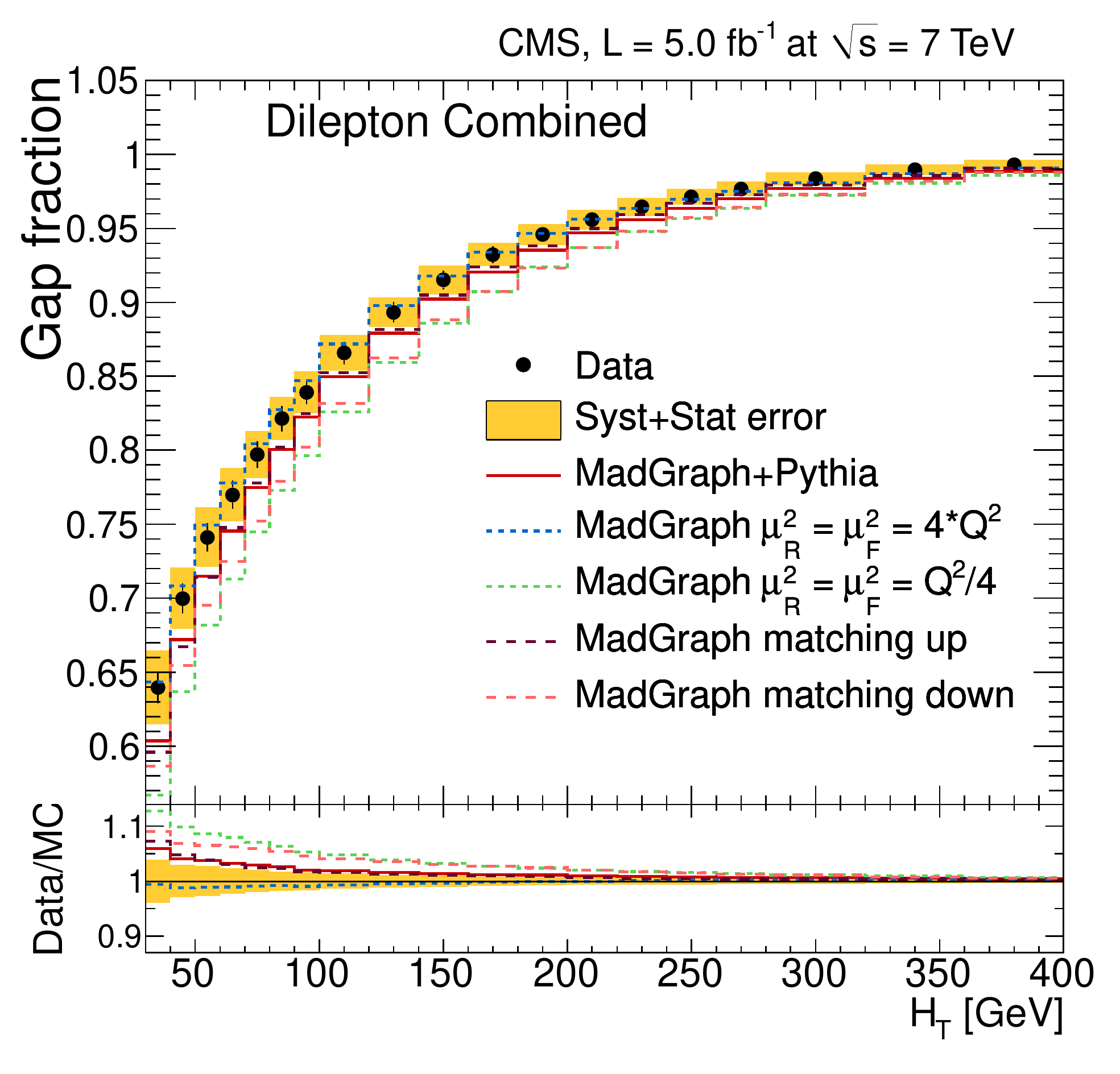}
\caption{Measured gap fraction as a function of the additional jet \pt (left) and of $\HT= \sum{\pt^{\text{add. jets}}}$ (right) in the dilepton channels. Data are compared to predictions from \MADGPYT, \POWPYT, and \MCNLOHER (top), as well as from \MADGRAPH with varied renormalisation and factorisation scales, and jet-parton matching threshold (bottom). The error bars on the data points indicate the statistical uncertainty. The shaded band corresponds to the combined statistical and total systematic uncertainty (added in quadrature).}
\label{fig:gap_dilep}
\end{figure*}

The measured gap-fraction distribution is compared to predictions from \MADGPYT, \POWPYT, and \MCNLOHER, and to the predictions from the \MADGRAPH samples with varied renormalisation and factorisation scales and jet-parton matching threshold. In Fig.~\ref{fig:gap_dilep} the gap fraction is measured as a function of the \pt of the leading additional jet (left) and as a function of $\HT$ (right), with the thresholds (defined at the abscissa where the data point is shown) varied between 35 and 380\GeV. The results are summarised in Table~\ref{tab:results:gap_pT} and Table~\ref{tab:results:gap_HT}, respectively. The measurements are consistent among the three dilepton channels. The gap fraction is lower as a function of $\HT$ showing that the measurement is probing quark and gluon emission beyond the first emission. The gap fraction is better described by \MCATNLO+\HERWIG compared to \MADGPYT and \POWPYT. This result is not incompatible with the observation described above, because the gap fraction requires the jets to have a certain \pt above the threshold, which does not imply necessarily large jet multiplicities. Decreasing the renormalisation and factorisation scales or matching threshold in the \MADGRAPH sample worsens the agreement between data and simulation.

\begin{table*}[htpb!]
\centering
\topcaption{Measured gap fraction as a function of the additional jet
\pt. The statistical, systematic, and total uncertainties are also shown. }
\label{tab:results:gap_pT}
\begin{tabular}{D{.}{.}{3.3}cccc}
\hline
\multicolumn{1}{c}{\pt Threshold (\GeVns{})} & Result & Stat. (\%) & Syst. (\%) & Total (\%) \\
\hline
 35	&	0.64	&	1.7	&	3.5	&	3.9	\\
 45	&	0.70	&	1.4	&	2.6	&	3.0	\\
 55	&	0.74	&	1.3	&	2.4	&	2.7	\\
 65	&	0.77	&	1.2	&	2.0	&	2.3	\\
 75	&	0.80	&	1.1	&	1.6	&	2.0	\\
 85	&	0.82	&	1.0	&	1.4	&	1.8	\\
 95	&	0.84	&	1.0	&	1.4	&	1.7	\\
110	&	0.87	&	0.9	&	1.1	&	1.4	\\
130	&	0.89	&	0.8	&	0.8	&	1.1	\\
150	&	0.92	&	0.7	&	0.8	&	1.1	\\
170	&	0.93	&	0.6	&	0.6	&	0.8	\\
190	&	0.95	&	0.6	&	0.5	&	0.7	\\
210	&	0.96	&	0.5	&	0.5	&	0.7	\\
230	&	0.96	&	0.4	&	0.5	&	0.6	\\
250	&	0.97	&	0.4	&	0.4	&	0.6	\\
270	&	0.98	&	0.4	&	0.4	&	0.5	\\
300	&	0.98	&	0.3	&	0.3	&	0.5	\\
340	&	0.99	&	0.3	&	0.3	&	0.4	\\
380	&	0.99	&	0.2	&	0.2	&	0.3	\\
\hline
\end{tabular}
\end{table*}

\begin{table*}[htpb!]
\centering
\topcaption{Measured gap fraction as a function of $\HT= \sum{\pt^{\text{add. jets}}}$. The statistical, systematic, and total uncertainties are also shown. }
\label{tab:results:gap_HT}
\begin{tabular}{D{.}{.}{3.3}cccc}
\hline
\multicolumn{1}{c}{$\HT$ Threshold (\GeVns{})} & Result & Stat. (\%) & Syst. (\%) & Total (\%) \\
\hline
 35	&	0.64	&	1.6	&	3.6	&	3.9	\\
 45	&	0.71	&	1.4	&	2.3	&	2.6	\\
 55	&	0.77	&	1.2	&	1.9	&	2.3	\\
 65	&	0.81	&	1.1	&	1.4	&	1.8	\\
 75	&	0.84	&	1.0	&	1.2	&	1.5	\\
 85	&	0.87	&	0.9	&	1.1	&	1.4	\\
 95	&	0.89	&	0.8	&	1.0	&	1.3	\\
110	&	0.91	&	0.7	&	0.8	&	1.1	\\
130	&	0.93	&	0.6	&	0.6	&	0.8	\\
150	&	0.95	&	0.5	&	0.6	&	0.8	\\
170	&	0.96	&	0.4	&	0.5	&	0.7	\\
190	&	0.97	&	0.4	&	0.4	&	0.6	\\
210	&	0.98	&	0.3	&	0.4	&	0.5	\\
230	&	0.98	&	0.3	&	0.3	&	0.4	\\
250	&	0.99	&	0.3	&	0.2	&	0.3	\\
270	&	0.99	&	0.2	&	0.2	&	0.3	\\
300	&	0.99	&	0.2	&	0.2	&	0.3	\\
340	&	1.00	&	0.2	&	0.2	&	0.2	\\
380	&	1.00	&	0.1	&	0.1	&	0.2	\\
\hline
\end{tabular}
\end{table*}

The total systematic uncertainty is about 3.5\% for values of the threshold (\pt or $\HT$) below 40\GeV, and decreases to 0.2\% for values of the thresholds above 200\GeV. Dominant sources of systematic uncertainty arise from the uncertainty in the JES and the background contamination, corresponding to approximately 2\% and 1\% systematic uncertainty, respectively, for the smallest \pt and $\HT$ values. Other sources with smaller impact on the total uncertainty are the b-tagging efficiency, JER, pileup, and the procedure used to correct the data to particle level.

\section{Summary}\label{sec:summary}

Measurements of the normalised differential \ttbar production cross section as a function of the number of jets in the dilepton (\Pe\Pe, $\Pgm\Pgm$, and $\Pe\mu$) and $\ell$+jets (\Pe+jets, $\mu$+jets) channels are presented. The measurements are performed using a data sample corresponding to an integrated luminosity of 5.0~\fbinv collected in pp~collisions at $\sqrt{s}= 7\TeV$ with the CMS detector. The results are presented in the visible phase space and compared with predictions of perturbative quantum chromodynamics from \MADGRAPH and \POWHEG interfaced with \PYTHIA, and \MCATNLO interfaced with \HERWIG, as well as \MADGRAPH with varied renormalisation and factorisation scales, and jet-parton matching threshold. The normalised differential \ttbar production cross section is also measured as a function of the jets radiated in addition to the \ttbar decay products in the $\ell$+jets channel.
The \MADGPYT and \POWPYT predictions describe the data well up to high jet multiplicities, while \MCNLOHER predicts fewer events with large number of jets.
The gap fraction is measured in dilepton events as a function of the \pt of the leading additional jet and the scalar sum of the \pt of the additional jets, and is also compared to different theoretical predictions. No significant deviations are observed between data and simulation.
The \MCNLOHER model seems to more accurately describe the gap fraction for all values of the thresholds compared to \MADGPYT and \POWPYT.

\section*{Acknowledgements}
{\tolerance=1200
We congratulate our colleagues in the CERN accelerator departments for the excellent performance of the LHC and thank the technical and administrative staffs at CERN and at other CMS institutes for their contributions to the success of the CMS effort. In addition, we gratefully acknowledge the computing centres and personnel of the Worldwide LHC Computing Grid for delivering so effectively the computing infrastructure essential to our analyses. Finally, we acknowledge the enduring support for the construction and operation of the LHC and the CMS detector provided by the following funding agencies: BMWFW and FWF (Austria); FNRS and FWO (Belgium); CNPq, CAPES, FAPERJ, and FAPESP (Brazil); MES (Bulgaria); CERN; CAS, MoST, and NSFC (China); COLCIENCIAS (Colombia); MSES and CSF (Croatia); RPF (Cyprus); MoER, SF0690030s09 and ERDF (Estonia); Academy of Finland, MEC, and HIP (Finland); CEA and CNRS/IN2P3 (France); BMBF, DFG, and HGF (Germany); GSRT (Greece); OTKA and NIH (Hungary); DAE and DST (India); IPM (Iran); SFI (Ireland); INFN (Italy); NRF and WCU (Republic of Korea); LAS (Lithuania); MOE and UM (Malaysia); CINVESTAV, CONACYT, SEP, and UASLP-FAI (Mexico); MBIE (New Zealand); PAEC (Pakistan); MSHE and NSC (Poland); FCT (Portugal); JINR (Dubna); MON, RosAtom, RAS and RFBR (Russia); MESTD (Serbia); SEIDI and CPAN (Spain); Swiss Funding Agencies (Switzerland); MST (Taipei); ThEPCenter, IPST, STAR and NSTDA (Thailand); TUBITAK and TAEK (Turkey); NASU and SFFR (Ukraine); STFC (United Kingdom); DOE and NSF (USA).

Individuals have received support from the Marie-Curie programme and the European Research Council and EPLANET (European Union); the Leventis Foundation; the A. P. Sloan Foundation; the Alexander von Humboldt Foundation; the Belgian Federal Science Policy Office; the Fonds pour la Formation \`a la Recherche dans l'Industrie et dans l'Agriculture (FRIA-Belgium); the Agentschap voor Innovatie door Wetenschap en Technologie (IWT-Belgium); the Ministry of Education, Youth and Sports (MEYS) of Czech Republic; the Council of Science and Industrial Research, India; the Compagnia di San Paolo (Torino); the HOMING PLUS programme of Foundation for Polish Science, cofinanced by EU, Regional Development Fund; and the Thalis and Aristeia programmes cofinanced by EU-ESF and the Greek NSRF.
\par}

\ifthenelse{\boolean{cms@external}}{\vspace*{4ex}}{}
\bibliography{auto_generated}
\cleardoublepage \appendix\section{The CMS Collaboration \label{app:collab}}\begin{sloppypar}\hyphenpenalty=5000\widowpenalty=500\clubpenalty=5000\textbf{Yerevan Physics Institute,  Yerevan,  Armenia}\\*[0pt]
S.~Chatrchyan, V.~Khachatryan, A.M.~Sirunyan, A.~Tumasyan
\vskip\cmsinstskip
\textbf{Institut f\"{u}r Hochenergiephysik der OeAW,  Wien,  Austria}\\*[0pt]
W.~Adam, T.~Bergauer, M.~Dragicevic, J.~Er\"{o}, C.~Fabjan\cmsAuthorMark{1}, M.~Friedl, R.~Fr\"{u}hwirth\cmsAuthorMark{1}, V.M.~Ghete, C.~Hartl, N.~H\"{o}rmann, J.~Hrubec, M.~Jeitler\cmsAuthorMark{1}, W.~Kiesenhofer, V.~Kn\"{u}nz, M.~Krammer\cmsAuthorMark{1}, I.~Kr\"{a}tschmer, D.~Liko, I.~Mikulec, D.~Rabady\cmsAuthorMark{2}, B.~Rahbaran, H.~Rohringer, R.~Sch\"{o}fbeck, J.~Strauss, A.~Taurok, W.~Treberer-Treberspurg, W.~Waltenberger, C.-E.~Wulz\cmsAuthorMark{1}
% ...
\vskip\cmsinstskip
% ...
1:~~Also at Vienna University of Technology, Vienna, Austria\\
2:~~Also at CERN, European Organization for Nuclear Research, Geneva, Switzerland\\
% ...

\end{sloppypar}
\end{document}